\documentclass[conference,compsoc]{IEEEtran}

\ifCLASSOPTIONcompsoc
  \usepackage[nocompress]{cite}
\else
  \usepackage{cite}
\fi

\ifCLASSINFOpdf
\else
\fi

\usepackage{xspace}
\usepackage{enumitem}
\usepackage{multirow}
\usepackage{amssymb}
\usepackage{array}
\usepackage{threeparttable}
\usepackage{booktabs}
\usepackage{pifont}
\usepackage{cite}
\usepackage{float}
\usepackage{graphicx}
\usepackage{wrapfig} 
\usepackage[table,xcdraw]{xcolor}
\usepackage[colorlinks=true, linkcolor=black, citecolor=black, urlcolor=black]{hyperref}
\usepackage[export]{adjustbox}
\usepackage{tikz}
\usepackage{amsmath}
\usepackage{amssymb}
\usepackage[most, breakable]{tcolorbox}
\usepackage{bbding}
\usepackage{listings}
\usepackage{MnSymbol}
\usepackage{xcolor}

\newcommand{\zmm}[1] {}
\newcommand{\xu}[1] {}
\newcommand{\zj}[1] {}

\definecolor{mygreen1}{rgb}{0,0.6,0}
\definecolor{mygray1}{rgb}{0.5,0.5,0.5}
\definecolor{mymauve1}{rgb}{0.58,0,0.82}
\newcommand{\dingnum}[1]{%
  \ifcase#1
    \or \ding{182} %
    \or \ding{183} %
    \or \ding{184} %
    \or \ding{185} %
    \or \ding{186} %
    \or \ding{187} %
    \or \ding{188} %
    \or \ding{189} %
    \or \ding{190} %
    \or \ding{191} %
  \fi
}

\newcommand{\increase}[1]{\textsubscript{\textcolor{green!60!black}{$\uparrow$ #1}}}

\newcommand{\tool}{VulnResolver\xspace}

\newcommand{\CodeSearchToolkit}{Code Search Toolkit\xspace}
\newcommand{\CodeSymbolAnalysisToolkit}{Code Symbol Analysis Toolkit\xspace}
\newcommand{\PoCExecutionToolkit}{PoC Execution Toolkit\xspace}
\newcommand{\ProjectEditingToolkit}{Project Editing Toolkit\xspace}
\newcommand{\PythonCodeExecutionToolkit}{Python Code Execution Toolkit\xspace}
\newcommand{\SearchCodeElementInFile}{search\_code\_element\xspace}
\newcommand{\SearchCodeElementInFileWithSig}{search\_code\_element(name, file, mark\_lines)\xspace}
\newcommand{\ReadCodeInFile}{read\_code\xspace}
\newcommand{\ReadCodeInFileWithSig}{read\_code(file, center, num, mark\_lines)\xspace}
\newcommand{\ResolveCodeSymbol}{resolve\_code\_symbol\xspace}
\newcommand{\ResolveCodeSymbolWithSig}{resolve\_code\_symbol(queries)\xspace}
\newcommand{\RunPoC}{run\_poc\xspace}
\newcommand{\RunPoCWithSig}{run\_poc(unique\_name)\xspace}
\newcommand{\ApplyEdits}{apply\_edits\xspace}
\newcommand{\ApplyEditsWithSig}{apply\_edits(unique\_name, edits)\xspace}
\newcommand{\RollbackTheLatestOneEdit}{rollback\_the\_latest\_one\_edit\_set\xspace}
\newcommand{\RollbackTheLatestOneEditWithSig}{rollback\_the\_latest\_one\_edit\_set()\xspace}
\newcommand{\RollbackAllAppliedEdits}{rollback\_all\_applied\_edits\xspace}
\newcommand{\RollbackAllAppliedEditsWithSig}{rollback\_all\_applied\_edits()\xspace}
\newcommand{\RunPythonCode}{run\_python\_code\xspace}
\newcommand{\RunPythonCodeWithSig}{run\_python\_code(code)\xspace}

\newcommand{\preCCAgent}{Context Pre-Collection Agent\xspace}
\newcommand{\shortPreCCAgent}{\textit{CPCAgent}\xspace}
\newcommand{\spaAgent}{Safety Property Analysis Agent\xspace}
\newcommand{\shortSPAAgent}{\textit{SPAAgent}\xspace}

\newcommand{\secBenchFull}{SEC-bench Full\xspace}
\newcommand{\secBenchFirstEighty}{SEC-bench Lite\xspace}

\begin{document}
\title{{\tool}: A Hybrid Agent Framework for LLM-Based Automated Vulnerability Issue Resolution}

\author{
Mingming Zhang$^{1}$,
Xu Wang$^{1}$,
Jian Zhang$^{1}$\IEEEauthorrefmark{1},
Xiangxin Meng$^{1}$,
Jiayi Zhang$^{2}$,
Chunming Hu$^{1}$\\

$^{1}$Beihang University,
$^{2}$Nanyang Technological University\\

\textit{\{mmzhang, xuwang, zhangj3353, mengxx, hucm\}@buaa.edu.cn}, \textit{jzhang150@e.ntu.edu.sg}
}

\maketitle
\makeatletter
\long\def\@makefntext#1{\parindent 1em\noindent #1}
\makeatother
\footnotetext{\IEEEauthorrefmark{1} Corresponding author}

\begin{abstract}

As software systems grow in complexity, security vulnerabilities have become increasingly prevalent, posing serious risks and economic costs. Although automated detection tools such as fuzzers have advanced considerably, effective resolution still often depends on human expertise. Existing automated vulnerability repair (AVR) methods rely heavily on manually provided annotations (e.g., fault locations or CWE labels), which are often difficult and time-consuming to obtain, while overlooking the rich, naturally embedded semantic context found in issue reports from developers.

In this paper, we present {\tool}, the first LLM-based hybrid agent framework for automated vulnerability issue resolution. {\tool} unites the adaptability of autonomous agents with the stability of workflow-guided repair through two specialized agents. The \preCCAgent (\shortPreCCAgent) adaptively explores the repository to gather dependency and contextual information, while the \spaAgent (\shortSPAAgent) generates and validates the safety properties violated by vulnerabilities. Together, these agents produce structured analyses that enrich the original issue reports, enabling more accurate vulnerability localization and patch generation.

Evaluations on the SEC-bench benchmark show that {\tool} resolves 75\% of issues on \secBenchFirstEighty, achieving the best resolution performance. On \secBenchFull, {\tool} also significantly outperforms the strongest baseline, the agent-based OpenHands, confirming its effectiveness. Overall, {\tool} delivers an adaptive and security-aware framework that advances end-to-end automated vulnerability issue resolution through workflow stability and the specialized agents’ capabilities in contextual reasoning and property-based analysis.

\end{abstract}

\IEEEpeerreviewmaketitle

\section{Introduction}
\label{sec:intro}

Real-world software projects are increasingly complex and interconnected, leading to a higher prevalence of security vulnerabilities~\cite{blogReviewZeroday, cve, yu2025patchagent}, which elevate security risks and impose substantial economic costs~\cite{anwar2020measuring, iannone2022secret, cybersecurityventuresCybercrimeCost}. Timely mitigation is therefore crucial for maintaining software security and reliability. Vulnerability detection techniques such as fuzzing have uncovered thousands of vulnerabilities in practice~\cite{serebryany2017oss, manes2019fuzzingSurvey}, and bug finders typically file issue reports to notify maintainers (e.g., NJS-482~\cite{githubIssueNJS482}).
Despite advances in automated detection, fixing these issues correctly and effectively still relies heavily on human expertise, motivating automated approaches for vulnerability fixing.

Automated Program Repair (APR) aims to fix software bugs automatically~\cite{liu2019tbar, le2012GenProg}, and LLM-based APR methods~\cite{xia2024chatRepair, yin2024thinkrepair, bouzenia2024repairagent, zhang2025ReinFix} have shown strong performance. The introduction of SWE-bench~\cite{jimenez2023swebench} further promotes fully automated issue resolution, inspiring SWE systems~\cite{yang2024sweAgent, wang2024openhands, gauthier2024aider, bytedance2025TRAE, liu2024marscode, xia2025agentless, ruan2025SpecRover, antoniades2025swesearchenhancingsoftwareagents} for real-world bug fixing.
In the context of vulnerabilities, several Automated Vulnerability Repair (AVR) methods~\cite{li2025sokAVR, fu2022vulrepair, zhou2024vulMaster, pearce2023examiningZeroShotVulRepairWithLLMs, yu2025patchagent, kim2025SAN2PATCH, chen2023vrepair, nong2025appatch} reduce manual effort and automate mitigation, but many (e.g., VRepair~\cite{chen2023vrepair}, VulRepair~\cite{fu2022vulrepair}, APPATCH~\cite{nong2025appatch}) still require human-provided fault locations or CWE labels. Approaches like PatchAgent~\cite{yu2025patchagent} and SAN2PATCH~\cite{kim2025SAN2PATCH} automate localization and patching from sanitizer logs but ignore semantic context in issue reports, which often contains crucial clues for understanding and fixing vulnerabilities.
This gap motivates systems that can directly resolve vulnerability issues using both structured and semantic insights.

However, effective vulnerability resolution remains challenging. \textbf{First}, current SWE paradigms can be broadly categorized into agent-based (e.g., SWE-agent~\cite{yang2024sweAgent}) and workflow-based (e.g., Agentless~\cite{xia2025agentless}) frameworks, each with distinct advantages and limitations. Agent-based systems leverage the ReAct paradigm~\cite{yao2023react} to iteratively invoke tools, with each LLM agent selecting subsequent actions based on prior reasoning and tool feedback. Despite their flexibility, they often suffer from non-deterministic planning and suboptimal exploration in a large action space, frequently producing incomplete or abandoned patches~\cite{xia2025agentless}. In contrast, workflow-based systems such as Agentless~\cite{xia2025agentless} use a deterministic pipeline for localization and repair, avoiding uncontrolled agent behaviors and achieving higher stability. Nevertheless, Agentless relies solely on issue reports for vulnerability localization and context construction, completely forgoing repository-level adaptive exploration, limiting its ability to gather sufficient contextual information for accurate repair. A promising direction, therefore, is to combine the strengths of both paradigms: a hybrid agent system that operates under a deterministic workflow while enabling targeted repository exploration by agents (\textbf{Insight~1}).

\textbf{Second}, existing SWE systems primarily address generic software issues and lack designs tailored to the unique characteristics of security vulnerabilities. Vulnerabilities often arise from violations of specific security constraints~\cite{propertyGPT, useSafePropToGenVulPatch}. For example, a CWE-125 (Out-of-Bounds Read) typically occurs when a bounded buffer is accessed beyond its valid index range, violating the constraint that indices must remain within bounds. Such constraints can be formalized as safety properties, which stipulate that ``something bad never happens'' during program execution~\cite{SafetyProperties}. Violations of these properties correspond to exploitable vulnerabilities. Although safety properties have been leveraged in vulnerability detection~\cite{propertyGPT}, they remain underexplored in both SWE systems and recent LLM-based AVR methods. For instance, while PatchAgent~\cite{yu2025patchagent} and SAN2PATCH~\cite{kim2025SAN2PATCH} focus on end-to-end vulnerability repair, they incorporate no design components that explicitly encode or leverage vulnerability-specific constraints. Incorporating safety property reasoning into vulnerability issue resolution is therefore crucial for enabling more principled, effective, and security-aware automated repair (\textbf{Insight~2}).

In this paper, we propose {\tool}, an LLM-based framework designed for automated vulnerability issue resolution. \textbf{1) Combining agent flexibility with workflow determinism}, {\tool} builds upon an Agentless-style issue resolution workflow~\cite{xia2025agentless} while integrating adaptive tool-calling capabilities through specialized agents, resulting in a hybrid agent-based automated issue resolution framework. The workflow-driven structure ensures stability and consistency across key repair stages including vulnerability localization, patch generation, and patch selection, while the agents enhance adaptability and contextual understanding through targeted repository exploration. By maintaining a deterministic workflow backbone, {\tool} avoids the uncontrolled decision drift in fully agent-based or multi-agent systems~\cite{yang2024sweAgent, wang2024openhands}. \textbf{2) Addressing adaptive context acquisition}, we introduce the \textit{\preCCAgent (\shortPreCCAgent)}, which leverages a suite of static analysis–based code search and symbol resolution tools to perform structured exploration of the target repository. Guided by the LLM's reasoning, {\shortPreCCAgent} incrementally gathers relevant contextual information until sufficient coverage is achieved, after which it synthesizes the collected context and derived insights into a comprehensive context analysis report (Report~I). This process endows the workflow with repository awareness and enhances subsequent localization and repair processes. \textbf{3) Capturing vulnerability characteristics}, {\tool} incorporates the \textit{\spaAgent (\shortSPAAgent)}, which employs both static analysis and dynamic execution tools to infer, generate, and validate safety properties that describe secure program behavior. Specifically, {\shortSPAAgent} explores the vulnerable codebase and proof-of-concept (PoC) executions to hypothesize potential safety properties, instrument the repository with these properties, and verify them through guided PoC re-execution. The agent then iteratively refines its property hypotheses based on validation results, achieving a deeper semantic understanding of the vulnerability's root cause. Once convergence is reached, SPAAgent generates a property analysis report (Report~II) that encapsulates generated properties and their security implications.

Finally, {\tool} integrates the issue report with the two agent-generated reports (Report~I and Report~II) into an enhanced issue report, consolidating insights from repository exploration and safety property reasoning. This enriched input improves localization and patch generation, unifying workflow determinism with agent reasoning for a principled, adaptive, and security-aware AVR system.

We conduct extensive experiments on SEC-bench~\cite{lee2025secBench} to evaluate the effectiveness of {\tool}. Following the evaluation protocol of its original paper, we construct two benchmark variants, \secBenchFirstEighty and \secBenchFull. {\tool} consistently and substantially outperforms all baselines, successfully resolving 75\% of issues on \secBenchFirstEighty and achieving the best performance among all compared methods. Furthermore, the complete {\tool} configuration achieves a 53.8\% improvement over the base workflow (75.0\% vs.\ 48.8\%), demonstrating the effectiveness of our overall design. Finally, generalization experiments on \secBenchFull show that {\tool} successfully resolves 67.5\% of issues, outperforming the best leaderboard baseline, OpenHands~\cite{wang2024openhands} + Claude-3.7-Sonnet~\cite{anthropicClaude37Sonnet}, by 98.5\% and our base workflow by 37.8\%, further confirming the effectiveness of our approach.
The main contributions of this paper are summarized as follows:

\begin{itemize}
	\item \textbf{Direction.} To our knowledge, {\tool} is the first to explore \emph{automated vulnerability issue resolution}, directly and autonomously addressing real-world vulnerability issues and establishing a new research direction for AVR. We further introduce the \emph{hybrid agent} paradigm, combining workflow stability with agent flexibility. Together, these two directions advance LLM-based software engineering toward more practical, autonomous, and security-aware program repair.

    \item \textbf{Technique.} We present {\tool}, an LLM-based hybrid agent system that integrates a deterministic workflow with two specialized agents: the \emph{\preCCAgent (\shortPreCCAgent)} for adaptive repository exploration and the \emph{\spaAgent (\shortSPAAgent)} for property-based vulnerability reasoning. By combining context acquisition with property-based semantic understanding, {\tool} constructs enhanced issue reports that drive more effective vulnerability localization and patch generation.

    \item \textbf{Extensive Study.} We conduct extensive experiments on SEC-bench~\cite{lee2025secBench}, the first standardized benchmark for this task. {\tool} outperforms all baselines, achieving a 53.8\% improvement over the base workflow (75.0\% vs.\ 48.8\%) on \secBenchFirstEighty. Further experiments on \secBenchFull demonstrate strong generalization, surpassing the best leaderboard system by a large margin. An ablation study further validates the contribution of each agent and the overall framework design, confirming the effectiveness of our approach.
\end{itemize}

\begin{figure*}[t]
  \centering
  \includegraphics[width=\textwidth]{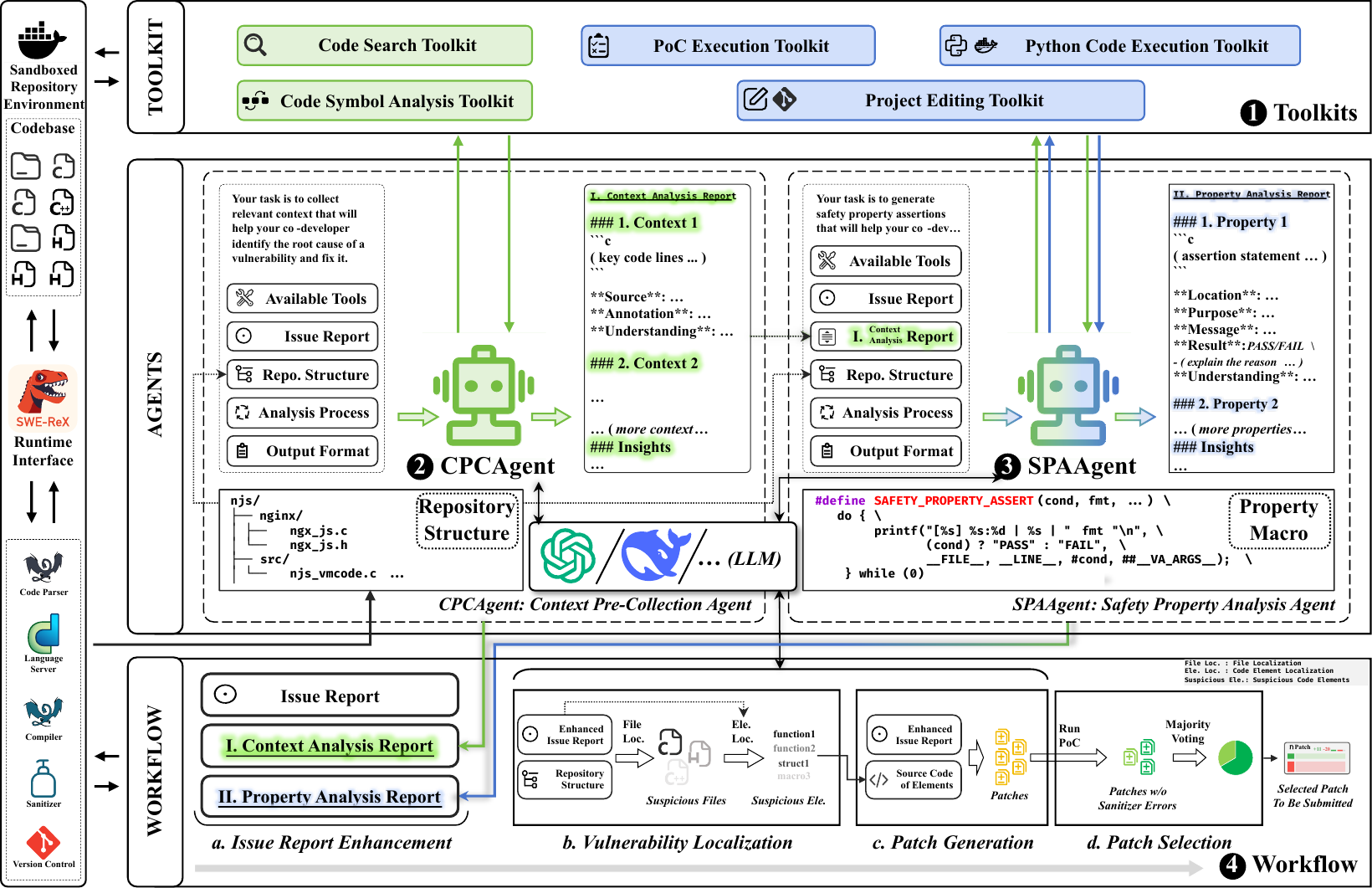}
  \caption{\textbf{An Overview of {\tool}.} Our approach integrates agents into a workflow. Two LLM-based agents are introduced, each achieving its objective with support from \dingnum{1}toolkits. \dingnum{2}{\shortPreCCAgent}, with {\CodeSearchToolkit} and {\CodeSymbolAnalysisToolkit}, adaptively collects context to generate the I. Context Analysis Report. \dingnum{3}{\shortSPAAgent}, supported by five toolkits, generates safety properties through assertion insertion, which are dynamically validated and iteratively refined, resulting in the II. Property Analysis Report. These reports, combined with the original issue report, form an enhanced issue report that drives our \dingnum{4}workflow. The LLM-based workflow resolves issues through vulnerability localization, patch generation, and patch selection.
  } 
  \label{fig:overview}
\end{figure*}

\section{Approach}
\label{sec:approach}

\subsection{Overview} 
\label{sec:approach_overview}

Now we introduce {\tool}, a hybrid LLM-based framework for automated vulnerability issue resolution that combines the adaptability of agent-based systems with the stability of workflow-driven repair. As illustrated in Figure~\ref{fig:overview}, {\tool} integrates two specialized agents, each equipped with \dingnum{1}toolkits (Section~\ref{sec:AgentToolkits}) operating within a sandboxed environment of the target repository. These agents, the Context Pre-Collection Agent (\dingnum{2}{\shortPreCCAgent}, Section~\ref{sec:preCCAgent}) and the Safety Property Analysis Agent (\dingnum{3}{\shortSPAAgent}, Section~\ref{sec:spaAgent}), work together to enhance vulnerability analysis and repair. Specifically, the {\shortPreCCAgent} adaptively explores the target codebase using static analysis tools to gather critical, vulnerability-related contextual information, resulting in a Context Analysis Report (Report I). In contrast, the {\shortSPAAgent} leverages both static analysis and dynamic execution tools to iteratively analyze, generate, insert, and validate safety properties that describe secure program behavior, ultimately producing a Property Analysis Report (Report II). When these two reports are combined with the original issue report, they form an enhanced issue report (\(\mathbf{a}\)). This enhanced report is then used as input for our \dingnum{4}base workflow (Section~\ref{sec:workflow}), which includes stages for vulnerability localization (\(\mathbf{b}\)), patch generation (\(\mathbf{c}\)), and patch selection (\(\mathbf{d}\)). By integrating the adaptive capabilities of the agents with a structured workflow, {\tool} forms an effective framework for real-world AVR.

\subsection{Agent Toolkits}
\label{sec:AgentToolkits}

\begin{table*}
\centering
\caption{Agent Toolkits Integrated into {\tool}}
\label{tab:agentToolkits}

\resizebox{\textwidth}{!}{
\begin{tabular}{l|l|cc}
\toprule
\textbf{Toolkit} & \textbf{Description} & \textbf{\shortPreCCAgent} & \textbf{\shortSPAAgent} \\ \hline

\multirow{1}{*}{\textbf{\CodeSearchToolkit (\S\ref{sec:codeSearchToolkit})}} & \multirow{1}{*}{Retrieve code and annotate \texttt{mark\_lines} with special markers.} & \multirow{3}{*}{\ding{51}} & \multirow{3}{*}{\ding{51}} \\
\quad \SearchCodeElementInFileWithSig & \quad Search code elements by \texttt{name} and return their implementation. & & \\
\quad \ReadCodeInFileWithSig & \quad Read \texttt{center}-th line with surrounding \texttt{num} lines. & & \\ \hline

\textbf{\CodeSymbolAnalysisToolkit (\S\ref{sec:codeSymbolAnalysisToolkit})} & \multirow{1}{*}{Resolve code symbols to identify dependencies.} & \multirow{2}{*}{\ding{51}} & \multirow{2}{*}{\ding{51}} \\
\quad \ResolveCodeSymbolWithSig & \quad Find definitions and references using marker-based \texttt{queries}. & & \\ \hline

\textbf{\PoCExecutionToolkit (\S\ref{sec:pocExecutionToolkit})} & \multirow{1}{*}{Execute the target PoC to check vulnerability behavior.} & & \multirow{2}{*}{\ding{51}} \\
\quad \RunPoCWithSig & \quad Compile and run the PoC program; and return execution output. & & \\ \hline

\multirow{1}{*}{\textbf{\ProjectEditingToolkit (\S\ref{sec:projectEditingToolkit})}} & \multirow{1}{*}{Apply and manage edits to insert or fix safety properties.} & \multirow{4}{*}{} & \multirow{4}{*}{\ding{51}} \\
\quad \ApplyEditsWithSig & \quad Apply \texttt{edits} and record them as a Git commit. & & \\
\quad \RollbackTheLatestOneEditWithSig & \quad Revert the most recent commit. & & \\
\quad \RollbackAllAppliedEditsWithSig & \quad Revert all applied edits. & & \\ \hline

\textbf{\PythonCodeExecutionToolkit (\S\ref{sec:pythonCodeExecutionToolkit})} & \multirow{1}{*}{Run Python code for PoC output and other lightweight analysis.} & & \multirow{2}{*}{\ding{51}} \\
\quad \RunPythonCodeWithSig & \quad Execute agent-generated Python code and return execution output. & & \\ \bottomrule

\end{tabular}
} %

\end{table*}

To support our  {\shortPreCCAgent} and {\shortSPAAgent}, we design a series of toolkits integrated into the agents as actions. These toolkits enable the agents to analyze vulnerabilities and accomplish their goals: context pre-collection and safety property generation.

As shown in Table~\ref{tab:agentToolkits}, we consider: (1) \textbf{\CodeSearchToolkit} for searching and reading source code in the target repository; (2) \textbf{\CodeSymbolAnalysisToolkit} for resolving code dependencies (e.g., finding function definitions); (3) \textbf{\PoCExecutionToolkit} for compiling the repository and running PoCs in a controlled environment; (4) \textbf{\ProjectEditingToolkit} for modifying the repository and inserting safety property assertions; and (5) \textbf{\PythonCodeExecutionToolkit}, enabling agents to run self-generated Python scripts for flexible computations such as post-processing PoC outputs, numeric checking, and other lightweight analyses.

These five toolkits form the operational foundation of the two proposed agents, facilitating automated understanding and reasoning over vulnerability issues to achieve their respective objectives.

\subsubsection{\CodeSearchToolkit}\label{sec:codeSearchToolkit} This toolkit aids agents in retrieving code in vulnerable repositories, consisting of two tools: (1) \textbf{\SearchCodeElementInFile}, which searches for code elements by name within a file and returns their implementation (e.g., handling seven types in C/C++: classes, structs, unions, enums, functions, macros, and global variables); and (2) \textbf{\ReadCodeInFile}, which retrieves code by line numbers, returning the target line and a configurable number of surrounding lines. Both tools provide the relevant code along with metadata, such as element name, element type, file path, and line number range. Existing APR studies~\cite{liu2024marscode, yu2025patchagent} mention that LLMs struggle with calculating line numbers. However, issue reports often include line numbers (e.g., in stack traces), which may confuse LLMs regarding which specific code is being referenced. To address this, we extend these tools with a parameter for specifying line numbers of interest. The corresponding code lines returned by both tools are annotated with markers. For example:
\begin{lstlisting}[language=c, firstnumber=149]
    ...
    for (i = 0; i < length; i++) {
        if (njs_is_valid(&array->start[i])) { // <<<<< njs/src/njs_array.c:151
            njs_uint32_to_string(&index, i);
            ...
\end{lstlisting}
This enables accurate mapping of line numbers to code, which is crucial for interpreting line-numbered descriptions in issue reports, such as crash locations.

\subsubsection{\CodeSymbolAnalysisToolkit}\label{sec:codeSymbolAnalysisToolkit} Based on the issue report, the agents first identify relevant starting points within the target repository and retrieves the corresponding code using {\CodeSearchToolkit}. To gain a deeper understanding, the agent also requires tools to resolve code dependencies. Therefore, we introduce {\CodeSymbolAnalysisToolkit}, which includes \textbf{{\ResolveCodeSymbol}} to resolve code symbols and identify their definitions and references. This tool aligns with the Language Server Protocol (LSP)~\cite{microsoftOfficialPageLSP}, commonly used in IDEs like Visual Studio Code. Previous LLM-based APR methods~\cite{liu2024marscode, yu2025patchagent} have adopted similar functionality, requiring a line number and symbol name as input, but LLMs often struggle with numeric calculations, such as line numbers. While fuzzy matching can partially mitigate this, it may lead to imprecise resolutions due to symbol name overlaps. To address this, we propose a marker-based resolution mechanism, where the agent inserts markers into the code to indicate the symbols it wants to resolve, reframing symbol identification as a code-editing task. These markers, which are macro-like invocations (e.g., \texttt{FIND\_DEFINITION(symbol)}, \texttt{FIND\_REFERENCES(symbol)}), enable the obtaining of symbol locations by parsing the pre- and post-edit differences. The toolkit then queries the LSP server and returns the results to the agent. Edits are applied virtually, without modifying the repository, using the \emph{SEARCH/REPLACE} format~\cite{gauthier2024aider, xia2025agentless, huang2025guirepair}, consisting of a file path, a search string, and a replacement string. Multiple markers and edits can be handled in a single call, enabling the simultaneous resolution of multiple symbols and resolution types.

\begin{wrapfigure}{r}{0.22\textwidth}
    \centering
    \includegraphics[width=\linewidth]{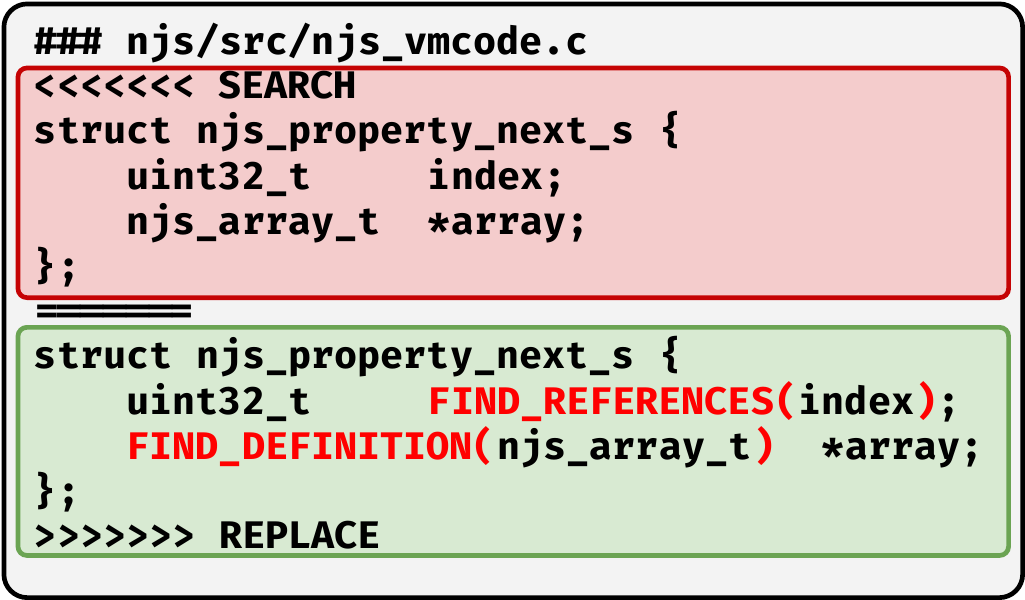}
    \caption{An Example Input to the \texttt{\ResolveCodeSymbol}}
    \label{fig:codeSymbolAnaExample}
\end{wrapfigure} For example, Figure~\ref{fig:codeSymbolAnaExample} illustrates a case where the agent queries the definition of \texttt{njs\_array\_t} and the references to \texttt{index} within a code snippet from njs/src/njs\_vmcode.c. \textbf{Conceptually}, the toolkit enables the LLM to emulate interactive IDE behaviors, such as selecting or clicking on code, effectively simulating LSP invocations like \texttt{Ctrl+Click}.

\subsubsection{\PoCExecutionToolkit}\label{sec:pocExecutionToolkit} 
This toolkit enables {\shortSPAAgent} to execute the PoC program for a given vulnerability and check if the inserted property assertions are triggered. It includes one tool, \textbf{{\RunPoC}}, which compiles the target repository, runs the PoC, and returns either compilation errors or execution logs, consisting of assertion outputs and sanitizer messages. These logs guide the agent in fixing compilation errors or refining properties. Since assertion statements are LLM-generated, execution counts are unbounded, and logs can be excessively long, potentially exceeding the LLM context window. To address this, we employ three strategies: (1) Truncate the logs, providing the shortened output. (2) Parse the logs to count passed and failed assertions, appending a summary at the top. (3) Store the full logs outside the LLM context window, in the toolkit's RAM memory, retrievable by executing self-generated code via {\PythonCodeExecutionToolkit} using \texttt{get\_poc\_output(name: str)}. Each {\RunPoC} call requires a unique name provided by the agent to retrieve the corresponding output using \texttt{get\_poc\_output}. If the name is not unique, the toolkit automatically appends a numeric suffix, which is a sequential count, to ensure uniqueness. Both the fixed name and the execution results are returned to the agent.

\subsubsection{\ProjectEditingToolkit}\label{sec:projectEditingToolkit} 
This toolkit enables {\shortSPAAgent} to insert property assertions by modifying files in the target repository. To assist the agent in fixing incorrect modifications, such as those causing compilation issues or redundant assertions, it records each set of edits as a Git commit, enabling rollback when necessary. It includes three tools: (1) \textbf{\ApplyEdits}, which applies a set of edits in the \emph{SEARCH/REPLACE} format, following Section~\ref{sec:codeSymbolAnalysisToolkit}; (2) \textbf{\RollbackTheLatestOneEdit}, which reverts the most recent commit, enabling recovery and further modifications; and (3) \textbf{\RollbackAllAppliedEdits}, which reverts all prior edits when the entire editing trajectory is deemed incorrect.

All tools return the execution status, failure reasons (e.g., no changes or empty rollback history), and the current commit history, including the number of commits, the full commit list, and the most recent commit. A meaningful and unique name is required for each \textbf{\ApplyEdits} call, which serves as the identifier in the toolkit's output.

\subsubsection{\PythonCodeExecutionToolkit}\label{sec:pythonCodeExecutionToolkit} This toolkit contains a single tool, {\RunPythonCode}, enabling {\shortSPAAgent} to execute self-generated Python code. As described in Section~\ref{sec:pocExecutionToolkit}, PoC outputs are truncated, with the full output stored outside the LLM context window. To allow the agent to analyze the complete output, we integrate a Python execution tool inspired by CodeAct~\cite{wang2024codeact}, useful for fine-grained reasoning and data manipulation. For example, the agent can use this tool to analyze long PoC outputs through string operations, regular expressions, and other built-in capabilities, and print results via \texttt{print(...)}, using our provided function \texttt{get\_poc\_output(name: str)} to retrieve PoC output by name. The agent can also perform other lightweight analyses, such as numerical inspection (e.g., converting index values to hexadecimal for overflow/underflow checks).

To prevent excessive outputs, Python results are truncated before being returned. Additionally, to mitigate risks of unsafe operations (e.g., file access or command execution), we adopt the following safeguards: (1) We prompt the LLM not to access the file system or execute shell commands via this tool; (2) Unsafe operations like file reading/writing and command execution are replaced with stubs raising exceptions when attempted; and (3) Python code runs in an isolated Docker container via \texttt{llm-sandbox}~\cite{githubGitHubVndeellmsandbox}, separate from the vulnerable repository instance. These measures ensure {\RunPythonCode} remains safe for lightweight analyses while enhancing the agent's capabilities.

\subsection{{\preCCAgent}}
\label{sec:preCCAgent}

Initially, we can simply adopt the agentless workflow~\cite{xia2025agentless} to resolve issues. However, this workflow lacks adaptive repository exploration. It relies solely on the limited information provided in issue reports to identify relevant files, and then performs fine-grained localization on skeletonized file representations where function bodies are omitted. This constrained view prevents the workflow from understanding essential code semantics and ultimately limits the effectiveness of the repair process.
To address these, we introduce the \preCCAgent (\shortPreCCAgent). As shown in Figure~\ref{fig:overview}, the agent performs static analysis of the target repository using {\CodeSearchToolkit} and {\CodeSymbolAnalysisToolkit}, as listed in Table~\ref{tab:agentToolkits}, producing a context analysis report that captures critical information and insights to support vulnerability diagnosis and repair. The overall structure for \shortPreCCAgent is illustrated in Figure~\ref{fig:overview}, with its components described below. 

\textbf{Objective.}
The objective of \shortPreCCAgent is to collect and organize contextual information that supports identifying the root cause of the target vulnerability. The agent analyzes the repository, retrieves relevant code snippets, and integrates them into a structured context analysis report.

\textbf{Available Tools.} This part details the tools available to the \shortPreCCAgent: (1) two tools from the code search toolkit (\S\ref{sec:codeSearchToolkit}) and (2) one tool from the code symbol analysis toolkit (\S\ref{sec:codeSymbolAnalysisToolkit}). The code search tools retrieve snippets, while the code symbol analysis toolkit resolves dependencies (e.g., function definitions) to gather additional context. Together, these tools enable thorough context collection.

\textbf{Issue Report \& Repository Structure.} These parts specify the issue-specific information: the issue report and a representation of the repository structure in a Linux \texttt{tree}-like format~\cite{xia2025agentless}. This representation allows the LLM to understand the repository organization and accurately reference source file paths. Following Agentless~\cite{xia2025agentless}, only source files (e.g., \texttt{.c}, \texttt{.h}, \texttt{.cpp}) are included to reduce noise.

\textbf{Analysis Process.} This part outlines the analysis process: \textbf{(1) Initial Analysis.} Examine the issue report to identify triggering conditions, vulnerability type, and crash location. \textbf{(2) Context Identification.} Determine which context elements (e.g., functions, variables, or lines surrounding the crash) should be collected. \textbf{(3) Context Collection.} Incrementally gather relevant code snippets using the tools, starting from the most directly relevant code and expanding until sufficient context is gathered. \textbf{(4) Report Generation.} Upon convergence, synthesize the collected context, and derived insights into a context analysis report.

\textbf{Output Format.} This part specifies the expected output (i.e., the context analysis report) format of the agent, as shown in Figure~\ref{fig:overview}. \textit{First}, the collected contexts should be enumerated, each including: (1) the source code; (2) its source (e.g., file, function name, and line number); (3) annotations linking the context to issue report (e.g., whether it appears in the stack trace and which frame it is in); and (4) an explanation of the rationale for collecting each context and its relevance to the vulnerability’s root cause. \textit{Second}, the agent should summarize the insights derived from the exploration process, providing a synthesized understanding of the vulnerability’s behavior.

\shortPreCCAgent is implemented using the ReAct~\cite{yao2023react} framework, iteratively invoking the toolkits, observing results, and refining its context collection until a coherent and informative report is produced.

\subsection{{\spaAgent}}
\label{sec:spaAgent}

A central challenge in automated vulnerability resolution is that most existing repair tools rely on syntactic cues or shallow heuristics and struggle to reveal the semantic conditions causing a vulnerability to manifest. Without understanding these conditions, the resulting repairs are often brittle and may not generalize beyond the observed symptom. To address this gap, we introduce the \spaAgent (\shortSPAAgent), designed to infer and formalize safety properties that characterize the intended safe behavior of the vulnerable code.

Safety properties specify conditions that prevent the program from reaching unintended or unsafe states during execution~\cite{SafetyProperties}. In the context of vulnerabilities, these properties constrain critical operations, and a violation directly corresponds to a defect. For instance, a buffer overflow is precisely the violation of a bounds-safety property governing array accesses.

To implement safety property reasoning, \shortSPAAgent expresses each inferred property through targeted program instrumentation. We introduce a \texttt{SAFETY\_PROPERTY\_ASSERT} macro (Figure~\ref{fig:overview}), which encodes a safety property as an assertion specifying its violation condition and an identifying message for tracking relevant program states. This design reduces property generation to inserting \texttt{SAFETY\_PROPERTY\_ASSERT} calls at semantically critical locations.

\shortSPAAgent integrates static analysis with dynamic execution of provided PoCs to infer, instrument, and validate these properties. By exploring the repository, analyzing failing executions, and iteratively refining its hypotheses, the agent produces a property analysis report that captures the semantic root cause of the vulnerability and provides actionable guidance for downstream repair. The overall structure of \shortSPAAgent is illustrated in Figure~\ref{fig:overview}.

\textbf{Objective.}
The goal of \shortSPAAgent is to generate safety property assertions that characterize secure program behavior. The agent analyzes the codebase and PoC execution traces to hypothesize candidate properties, instruments them into the repository, and refines them based on validation results.

\textbf{Available Tools.} This part details the toolkits available to the \shortSPAAgent: (1) the code search toolkit (\S\ref{sec:codeSearchToolkit}), (2) the code symbol analysis toolkit (\S\ref{sec:codeSymbolAnalysisToolkit}), (3) the PoC execution toolkit (\S\ref{sec:pocExecutionToolkit}), (4) the project editing toolkit (\S\ref{sec:projectEditingToolkit}), and (5) the Python code execution toolkit (\S\ref{sec:pythonCodeExecutionToolkit}). The code search and symbol analysis toolkits support static exploration, while PoC execution and Python execution enable dynamic verification. The project editing toolkit facilitates the instrumentation and iterative refinement of assertions.

\textbf{Issue Report \& Context Analysis Report \& Repository Structure.} These parts specify the issue-specific information, including the issue report and repository structure, similar to \shortPreCCAgent. Additionally, the context analysis report generated  by {\shortPreCCAgent} is also provided.

\textbf{Analysis Process.} This part outlines the workflow for iterative safety property generation: \textbf{(1) Initial PoC Execution.} Run the PoC without modification to observe the crash point or vulnerability manifestation. \textbf{(2) Property Hypothesis.} Analyze the crash point and surrounding code to generate initial safety property assertions (e.g., index bounds, pointer non-null checks) and insert them into the repository via the project editing toolkit. \textbf{(3) Property Validation.} Execute the PoC with the inserted assertions to check for violations. Evaluate the relevance of triggered or passed assertions, retaining those informative for understanding the vulnerability. \textbf{(4) Iterative Refinement.} Refine property assertions based on PoC execution results, analyzing backwards from crash points or unsuccessful assertions to generate more precise and sufficient safety properties. \textbf{(5) Report Generation.} Upon convergence, synthesize the generated assertions, validation results, and derived insights into a property analysis report.

\textbf{Output Format.} This part specifies the expected output (i.e., the property analysis report) format of the agent, as shwon in Figure~\ref{fig:overview}. \textit{First}, list all generated safety properies with: (1) the assertion statement and contextually relevant code lines, (2) code location, (3) purpose of the assertion, (4) execution result (PASS/FAIL), and (5) interpretation of the assertion and results. \textit{Second}, summarize insights regarding the vulnerability's behavior and root cause.

Similar to \shortPreCCAgent, \shortSPAAgent is implemented using the ReAct~\cite{yao2023react} framework, leveraging the available toolkits to iteratively generate, instrument, and refine safety properties until a stable and informative explanation of the vulnerability is achieved.

\subsection{Issue Resolution Workflow}
\label{sec:workflow}

As shown in Figure~\ref{fig:overview}, our workflow consists of four stages: (\(\mathbf{a}\)) \textbf{Report Enhancement}, where the agents {\shortPreCCAgent} (\S\ref{sec:preCCAgent}) and {\shortSPAAgent} (\S\ref{sec:spaAgent}) generate a \emph{Context Analysis Report} and a \emph{Property Analysis Report}, respectively, to enrich the original issue report; (\(\mathbf{b}\)) \textbf{Vulnerability Localization}, which identifies suspicious code elements based on the enhanced report; (\(\mathbf{c}\)) \textbf{Patch Generation}, which generates patch candidates using the enhanced report and localized elements; and (\(\mathbf{d}\)) \textbf{Patch Selection}, which selects the most appropriate patch from the candidates for final submission.

The latter three stages are inspired by Agentless~\cite{xia2025agentless}, a recent LLM-based workflow for automated issue resolution. By integrating these stages, our workflow combines the controllability of workflow-based systems with the flexibility of agent-based reasoning, enabling more effective and reliable automated vulnerability issue resolution.

\subsubsection{Report Enhancement} In this stage, {\tool} invokes {\shortPreCCAgent} and {\shortSPAAgent} to obtain a \emph{Context Analysis Report} and a \emph{Property Analysis Report}, respectively. These two reports are then concatenated with the original issue report to construct an enhanced issue report for subsequent vulnerability localization and patch generation.

\subsubsection{Vulnerability Localization}\label{sec:vulLoc} {\tool} identifies suspicious code elements through a two-step process: file localization and code element localization.

\textbf{File Localization.} Following Agentless~\cite{xia2025agentless}, we first scan the target repository and construct a concise Linux tree-like representation of its file and directory structure, as in Section~\ref{sec:preCCAgent}. This helps the LLM understand the repository structure and code file paths.
First, {\tool} provides the repository structure and enhanced issue report to the LLM, prompting it to identify the top $N$ suspicious files relevant to the vulnerability. Second, to complement the prompting-based localization, following Agentless~\cite{xia2025agentless}, {\tool} applies lightweight retrieval-based localization using embeddings. Rather than embedding all files, we first instruct the LLM to identify irrelevant folders to ignore, based on the repository structure and enhanced report. The remaining files are chunked, and embeddings are computed for each chunk. The enhanced issue report is embedded as a query, and cosine similarity ranks the chunks. The top $N$ files are retrieved. Finally, {\tool} combines two sets to produce the final list of suspicious files.

\textbf{Code Element Localization.} Given the localized suspicious files, this step identifies code elements that may require modification. For example, in C/C++, we consider classes, structs, unions, enums, functions, macros, and global variables, which collectively cover most code semantics. Due to the infeasibility of providing complete code context in real-world complex projects, {\tool} constructs a skeleton representation, replacing function bodies with ellipses (``\texttt{...}''), as in Agentless~\cite{xia2025agentless}. These skeletons, along with the enhanced issue report, are provided to the LLM for localization. The LLM returns a JSON list with code elements identified by file path and either a simple or multi-part identifier (e.g., \texttt{open\_file} for a global function, \texttt{File::open} for a C++ method). After filtering out nonexistent elements, {\tool} generates the final set of suspicious code elements.

\subsubsection{Patch Generation} Based on the localized suspicious code elements and the enhanced issue report, {\tool} performs patch generation. First, it extracts the full implementation of the suspicious elements from the repository. Next, these code snippets are concatenated to form a context prompt for the LLM. In addition, following Agentless~\cite{xia2025agentless}, the context of each snippet is further extended by including $M$ lines of code before and after the target element. {\tool} then instructs the LLM to generate \emph{SEARCH/REPLACE} patches as described in Section~\ref{sec:codeSymbolAnalysisToolkit}. For each issue, the LLM generates $T$ patch candidates.

\subsubsection{Patch Selection}\label{sec:patchSelection} 
Consistent with prior issue resolution studies~\cite{xia2025agentless, huang2025guirepair, ruan2025SpecRover, yang2024sweAgent, lee2025secBench}, only one patch is selected for submission. For each candidate, {\tool} applies it to the repository, compiles the PoC, and executes it. Patches not triggering sanitizer errors are retained. These patches are then normalized to remove superficial differences. Each modified file is processed with a code formatter and comment removal tool. Following existing systems~\cite{xia2025agentless, bytedance2025TRAE}, the final patch is selected using majority voting over the normalized patched repository, with the most frequent patch chosen for submission. The selected patch is then converted to Git diff format~\cite{gitscmDiffformatDocumentation}.

\section{Experimental Setup}

\subsection{Research Questions}

\newcommand{\firstRQ}{How does {\tool} compare with existing issue resolution methods?}
\newcommand{\shortFirstRQ}{Effectiveness Comparison}

\newcommand{\secondRQ}{How do different design choices affect {\tool}’s effectiveness and LLM API cost?}
\newcommand{\shortSecondRQ}{Ablation Study}

\newcommand{\thirdRQ}{How does {\tool} generalize to resolve additional vulnerability issues?}
\newcommand{\shortThirdRQ}{Generalizability Study}

\begin{enumerate}[leftmargin=*, label=\textbf{RQ\arabic*}]
    \item {\firstRQ} (\textbf{\shortFirstRQ})
    \item {\secondRQ} (\textbf{\shortSecondRQ})
    \item {\thirdRQ} (\textbf{\shortThirdRQ})
\end{enumerate}

\subsection{Benchmark}

We evaluate {\tool} on SEC-bench~\cite{lee2025secBench}, a benchmark derived from real-world vulnerability issues in open-source C/C++ repositories. It is introduced to evaluate LLM-based agents on tasks software security tasks, including PoC generation and vulnerability patching. We use it for the patching task, where APR systems generate secure patches based on an issue report, codebase, and working PoC. For each issue, the benchmark provides a Docker image with a \texttt{secb} command that compiles and executes the PoC, triggering sanitizer errors before the fix. SEC-bench also offers an automated tool to verify correctness of the submitted patch. We use this tool to evaluate our method and baselines.

The benchmark contains 200 vulnerabilities from 29 C/C++ projects (e.g., gpac, ImageMagick, mruby, libredwg, njs) spanning 16 CWEs (e.g., CWE-125, CWE-787, CWE-476, CWE-416, CWE-119). 
In alignment with the SEC-bench paper~\cite{lee2025secBench}, 
we also use two SEC-bench versions:  
\begin{itemize}
    \item \textbf{\secBenchFirstEighty}: the 80-issue subset for evaluating effectiveness (RQ1, RQ2);  
    \item \textbf{\secBenchFull}: the full 200-issue dataset for evaluating generalizability (RQ3).
\end{itemize}

\begin{table}[H]
\centering
\caption{Statistics of the benchmark datasets}
\label{tab:secbenchDetails}
\resizebox{\linewidth}{!}{
\begin{tabular}{lccccc}
\toprule
\multicolumn{1}{c}{\multirow{2}{*}{\textbf{Dataset}}} &
  \multirow{2}{*}{\textbf{\#Project}} &
  \multirow{2}{*}{\textbf{\#Issue}} &
  \multicolumn{3}{c}{\textbf{Patch Edit Statistics}} \\
\multicolumn{1}{c}{}                &    &     & \textbf{Avg. Files} & \textbf{Avg. Hunks} & \textbf{Avg. Lines} \\
\midrule
\secBenchFirstEighty & 13 & 80  & 1.10                & 1.59                & 11.03               \\
\secBenchFull        & 29 & 200 & 1.28                & 2.43                & 17.29 \\
\bottomrule
\end{tabular}
} %
\end{table}

Table~\ref{tab:secbenchDetails} summarizes the characteristics of both datasets, including the number of projects, the number of issues, and patch-level edit statistics such as the average number of files, hunks, and lines modified per issue.

\subsection{Implementation}
\label{sec:impl}

We implement {\tool} in Python, using libclang~\cite{llvmLibclangTutorial} for parsing C/C++ code elements, which is employed in both the Code Search Toolkit (\S\ref{sec:codeSearchToolkit}) and the element localization phase (\S\ref{sec:vulLoc}). For symbol analysis, including ``find definition" and ``find references" queries, we leverage clangd~\cite{llvmWhatClangd}, an LSP server for C/C++. We use SWE-ReX~\cite{githubGitHubSWEagentSWEReX}, a sandboxed framework for AI agents, to interact with Docker containers created from SEC-bench's issue images, enabling secure repository exploration, file modification, and PoC execution. Additionally, to execute code securely in an isolated environment, we use llm\_sandbox~\cite{githubGitHubVndeellmsandbox}, which runs Python code in dedicated Docker containers, supporting the Python Code Execution Toolkit (\S\ref{sec:pythonCodeExecutionToolkit}). We construct two agents, {\shortPreCCAgent} and {\shortSPAAgent}, using langchain~\cite{langchainLangChain}. We adopt three state-of-the-art LLMs as the base models in our experiments, including one open-source LLM, DeepSeek-V3.2-Exp~\cite{liu2024deepseekv3, deepseekIntroducingDeepSeekV32Exp}, and two closed-source LLMs, o3-mini~\cite{openaiOpenAIO3mini} and GPT-4o~\cite{openaiHelloGPT4o}, both of which were evaluated in the SEC-bench paper~\cite{lee2025secBench}. DeepSeek-V3.2-Exp is selected as our default base model, as it demonstrates strong performance on SWE tasks~\cite{deepseekIntroducingDeepSeekV32Exp} and offers high cost efficiency~\cite{githubDeepSeekV32ExpDeepSeek_V3_2pdfMain}.
For retrieval-based suspicious file localization, we employ OpenAI’s text-embedding-3-small~\cite{openaiTextEmbedding3Models}, accessed via remote API calls, following prior studies on issue resolution~\cite{xia2025agentless, huang2025guirepair}. Our experimental setup mainly follows the configuration of Agentless~\cite{xia2025agentless}: (1) By default, we query the LLM using greedy decoding (temperature = 0); (2) For file localization, we identify $N$=3 suspicious files based on prompting and embeddings computed with a chunk size of 512 and no overlap; (3) A context window of ±10 lines around each code element is used when constructing the context for patch generation. We generate $T$=5 patches per issue to balance patch diversity and API cost, and further explore the impact of this parameter in RQ2 (\S\ref{sec:rq2ImpactOfT}), where the first patch is generated using greedy decoding (temperature = 0) and the remaining patches are generated using sampling decoding (temperature = 1), following existing SWE practices~\cite{xia2025agentless, huang2025guirepair}. Finally, we use clang-format~\cite{llvmClangFormatx2014} and scc~\cite{githubGitHubJlefflersccsnapshots} to standardize the generated patches, as described in Section~\ref{sec:patchSelection}. We evaluate the performance of our method and baselines by directly using the automated evaluation scripts provided by SEC-bench.

\subsection{Baseline}

We consider three state-of-the-art SWE systems, SWE-agent~\cite{yang2024sweAgent}, OpenHands~\cite{wang2024openhands}, and Aider~\cite{gauthier2024aider}, evaluated on \secBenchFirstEighty using three LLMs reported in the SEC-bench paper (Claude 3.7 Sonnet~\cite{anthropicClaude37Sonnet}, GPT-4o~\cite{openaiHelloGPT4o}, and o3-mini~\cite{openaiOpenAIO3mini}), resulting in a total of nine baseline configurations. For these baselines, we directly adopt the results reported in the SEC-bench paper~\cite{lee2025secBench}. In addition, as described in Section~\ref{sec:impl}, we select DeepSeek-V3.2-Exp~\cite{deepseekIntroducingDeepSeekV32Exp} as {\tool}'s base model. To ensure a fair comparison, we re-evaluate the three SWE systems under this model on \secBenchFirstEighty, using the original SEC-bench-provided scripts~\cite{githubsecbenchSWEagent, githubsecbenchOpenHands, githubsecbenchAider} without any modifications beyond replacing the model. Moreover, PatchAgent~\cite{yu2025patchagent} also employs an agent-based approach and focuses on end-to-end vulnerability repair tasks. We evaluate it on {\secBenchFirstEighty} under DeepSeek-V3.2-Exp and use it as an additional baseline. Overall, this results in 13 baselines, consisting of four agent-based methods across four LLMs.

\subsection{Evaluation Metric}

Following existing issue resolution studies~\cite{lee2025secBench, xia2025agentless, huang2025guirepair}, we report (1) \textbf{\#Resolved} and \textbf{\%Resolved}: the number and proportion of vulnerability issues successfully resolved by the tool; and (2) \textbf{Avg.\$}: the average LLM API cost incurred when running the tool on each issue.

\begin{table}
\centering
\caption{Issue resolution performance on \secBenchFirstEighty. Here \#R(\%R) denotes \#Resolved (\%Resolved).}
\label{tab:rq1Main}
\resizebox{\linewidth}{!}{
\begin{tabular}{c|l|c|c}
\toprule
 &                   &                                     &        \\
\multirow{-2}{*}{\textbf{Method}} &
  \multirow{-2}{*}{\textbf{Model}} &
  \multirow{-2}{*}{\textbf{\#R(\%R)}} &
  \multirow{-2}{*}{\textbf{Avg.\$}} \\
\midrule
 & Claude 3.7 Sonnet & \cellcolor[HTML]{C6E7D1}27 (33.8\%) & \$1.29 \\
 & OpenAI GPT-4o     & \cellcolor[HTML]{D9EEE0}21 (26.2\%) & \$0.48 \\
 & OpenAI o3-mini    & \cellcolor[HTML]{CDE9D6}25 (31.2\%) & \$0.13 \\ 
\multirow{-4}{*}{SWE-agent~\cite{yang2024sweAgent}} &
  DeepSeek-V3.2-Exp &
  \cellcolor[HTML]{BDE3C9}30 (37.5\%) &
  \$0.11 \\
\midrule
 & Claude 3.7 Sonnet & \cellcolor[HTML]{CDE9D6}25 (31.2\%) & \$0.61 \\
 & OpenAI GPT-4o     & \cellcolor[HTML]{F3F9F8}12 (15.0\%) & \$1.53 \\
 & OpenAI o3-mini    & \cellcolor[HTML]{F9FBFD}10 (12.5\%) & \$0.15 \\
\multirow{-4}{*}{OpenHands~\cite{wang2024openhands}} &
  DeepSeek-V3.2-Exp &
  \cellcolor[HTML]{E7F4ED}16 (20.0\%) &
  \$0.05 \\
\midrule
 & Claude 3.7 Sonnet & \cellcolor[HTML]{E7F4ED}16 (20.0\%) & \$0.44 \\
 & OpenAI GPT-4o     & \cellcolor[HTML]{FCFCFF}9 (11.2\%)  & \$0.29 \\
 & OpenAI o3-mini    & \cellcolor[HTML]{EDF6F2}14 (17.5\%) & \$0.15 \\
\multirow{-4}{*}{Aider~\cite{gauthier2024aider}} &
  DeepSeek-V3.2-Exp &
  \cellcolor[HTML]{E7F4ED}16 (20.0\%) &
  \$0.04 \\
\midrule
PatchAgent~\cite{yu2025patchagent} &
  DeepSeek-V3.2-Exp &
  \cellcolor[HTML]{8DD0A0}46 (57.5\%) &
  \$0.10 \\
\midrule
\midrule
 & OpenAI GPT-4o     & \cellcolor[HTML]{9CD6AD}41 (51.3\%) & \$0.85 \\
 & OpenAI o3-mini    & \cellcolor[HTML]{8DD0A0}46 (57.5\%) & \$0.33 \\
\multirow{-3}{*}{\textbf{VulnResolver}} &
  DeepSeek-V3.2-Exp &
  \cellcolor[HTML]{63BE7B}\textbf{60 (75.0\%)} &
  \$0.07 \\
\bottomrule
\end{tabular}
}
\end{table}

\begin{table*}
\centering
\caption{Issue resolution performance per repository on \secBenchFirstEighty for different methods, where each cell shows the number of issues resolved (X) and the corresponding percentage (Y\%), with cell shading indicating higher effectiveness.}
\label{tab:rq1ByRepo}
\begin{tabular}{rc|c||cccc}
\toprule
 &
   &
  \multicolumn{5}{c}{\textbf{DeepSeek-V3.2-Exp}} \\
\multirow{-2}{*}{\textbf{Repository}} &
  \multirow{-2}{*}{\textbf{\#Issue}} &
  \textbf{{\tool}} &
  \textbf{PatchAgent} &
  \textbf{SWE-agent} &
  \textbf{OpenHands} &
  \textbf{Aider} \\
\midrule
exiv2 &
  1 &
  \cellcolor[HTML]{63BE7B}1(100.0\%) &
  \cellcolor[HTML]{FCFCFF}0(0.0\%) &
  \cellcolor[HTML]{63BE7B}1(100.0\%) &
  \cellcolor[HTML]{63BE7B}1(100.0\%) &
  \cellcolor[HTML]{63BE7B}1(100.0\%) \\
faad2 &
  11 &
  \cellcolor[HTML]{63BE7B}8(72.7\%) &
  \cellcolor[HTML]{8ACE9C}6(54.5\%) &
  \cellcolor[HTML]{D6EDDE}2(18.2\%) &
  \cellcolor[HTML]{E9F5EF}1(9.1\%) &
  \cellcolor[HTML]{FCFCFF}0(0.0\%) \\
gpac &
  26 &
  \cellcolor[HTML]{63BE7B}19(73.1\%) &
  \cellcolor[HTML]{74C589}17(65.4\%) &
  \cellcolor[HTML]{A4D9B3}11(42.3\%) &
  \cellcolor[HTML]{FCFCFF}0(0.0\%) &
  \cellcolor[HTML]{D4ECDD}5(19.2\%) \\
imagemagick &
  4 &
  \cellcolor[HTML]{63BE7B}4(100.0\%) &
  \cellcolor[HTML]{8ACE9C}3(75.0\%) &
  \cellcolor[HTML]{D6EDDE}1(25.0\%) &
  \cellcolor[HTML]{B0DDBD}2(50.0\%) &
  \cellcolor[HTML]{FCFCFF}0(0.0\%) \\
libarchive &
  2 &
  \cellcolor[HTML]{63BE7B}2(100.0\%) &
  \cellcolor[HTML]{B0DDBD}1(50.0\%) &
  \cellcolor[HTML]{FCFCFF}0(0.0\%) &
  \cellcolor[HTML]{FCFCFF}0(0.0\%) &
  \cellcolor[HTML]{B0DDBD}1(50.0\%) \\
libdwarf-code &
  1 &
  \cellcolor[HTML]{63BE7B}1(100.0\%) &
  \cellcolor[HTML]{FCFCFF}0(0.0\%) &
  \cellcolor[HTML]{FCFCFF}0(0.0\%) &
  \cellcolor[HTML]{FCFCFF}0(0.0\%) &
  \cellcolor[HTML]{FCFCFF}0(0.0\%) \\
libiec61850 &
  1 &
  \cellcolor[HTML]{63BE7B}1(100.0\%) &
  \cellcolor[HTML]{63BE7B}1(100.0\%) &
  \cellcolor[HTML]{63BE7B}1(100.0\%) &
  \cellcolor[HTML]{63BE7B}1(100.0\%) &
  \cellcolor[HTML]{FCFCFF}0(0.0\%) \\
libredwg &
  7 &
  \cellcolor[HTML]{A5D9B4}4(57.1\%) &
  \cellcolor[HTML]{63BE7B}7(100.0\%) &
  \cellcolor[HTML]{BBE2C7}3(42.9\%) &
  \cellcolor[HTML]{BBE2C7}3(42.9\%) &
  \cellcolor[HTML]{FCFCFF}0(0.0\%) \\
libsndfile &
  1 &
  \cellcolor[HTML]{63BE7B}1(100.0\%) &
  \cellcolor[HTML]{63BE7B}1(100.0\%) &
  \cellcolor[HTML]{FCFCFF}0(0.0\%) &
  \cellcolor[HTML]{FCFCFF}0(0.0\%) &
  \cellcolor[HTML]{63BE7B}1(100.0\%) \\
mruby &
  12 &
  \cellcolor[HTML]{63BE7B}10(83.3\%) &
  \cellcolor[HTML]{FCFCFF}4(33.3\%) &
  \cellcolor[HTML]{E3F2E9}5(41.7\%) &
  \cellcolor[HTML]{E3F2E9}5(41.7\%) &
  \cellcolor[HTML]{FCFCFF}4(33.3\%) \\
njs &
  11 &
  \cellcolor[HTML]{63BE7B}6(54.5\%) &
  \cellcolor[HTML]{B0DDBD}4(36.4\%) &
  \cellcolor[HTML]{8ACE9C}5(45.5\%) &
  \cellcolor[HTML]{FCFCFF}2(18.2\%) &
  \cellcolor[HTML]{D6EDDE}3(27.3\%) \\
openjpeg &
  2 &
  \cellcolor[HTML]{63BE7B}2(100.0\%) &
  \cellcolor[HTML]{B0DDBD}1(50.0\%) &
  \cellcolor[HTML]{FCFCFF}0(0.0\%) &
  \cellcolor[HTML]{FCFCFF}0(0.0\%) &
  \cellcolor[HTML]{B0DDBD}1(50.0\%) \\
wabt &
  1 &
  \cellcolor[HTML]{63BE7B}1(100.0\%) &
  \cellcolor[HTML]{63BE7B}1(100.0\%) &
  \cellcolor[HTML]{63BE7B}1(100.0\%) &
  \cellcolor[HTML]{63BE7B}1(100.0\%) &
  \cellcolor[HTML]{FCFCFF}0(0.0\%) \\
\midrule
\multicolumn{2}{c|}{\textbf{\#Resolved}} &
  \cellcolor[HTML]{63BE7B}60 &
  \cellcolor[HTML]{94D2A5}46 &
  \cellcolor[HTML]{CCE9D5}30 &
  \cellcolor[HTML]{FCFCFF}16 &
  \cellcolor[HTML]{FCFCFF}16 \\
\multicolumn{2}{c|}{\textbf{\%Resolved}} &
  \cellcolor[HTML]{63BE7B}75.0\% &
  \cellcolor[HTML]{94D2A6}57.5\% &
  \cellcolor[HTML]{CCE9D6}37.5\% &
  \cellcolor[HTML]{FCFCFF}20.0\% &
  \cellcolor[HTML]{FCFCFF}20.0\% \\
\bottomrule
\end{tabular}
\end{table*}

\section{Evaluation}

\subsection{RQ1: {\shortFirstRQ}}\label{sec:rq1}

To answer this RQ, we evaluate the effectiveness of {\tool} against 12 SWE methods (3 SWE systems $\times$ 4 base models) and a state-of-the-art AVR method, PatchAgent~\cite{yu2025patchagent}, on the {\secBenchFirstEighty} benchmark.

\textbf{Overall Performance.} The overall resolution rates on {\secBenchFirstEighty} are presented in Table~\ref{tab:rq1Main}. {\tool} successfully resolves 75.0\% of issues, outperforming the best SWE system, SWE-agent with DeepSeek-V3.2-Exp, which resolves 37.5\% of issues, by a margin of 100\%. {\tool} outperforming the best baseline, PatchAgent~\cite{yu2025patchagent} by 30.43\%, thus achieving the best performance across all evaluated methods. Using other LLMs, our method also demonstrates good performance. Specifically, {\tool} with o3-mini fixes 57.5\% of issues, and {\tool} with GPT-4o fixes 51.3\% of issues, both significantly outperforming the baselines under the same base models.

\textbf{Per-Project Performance.} We report the resolve rates of {\tool} and the baselines for each project in {\secBenchFirstEighty}. Notably, due to SEC-bench not reporting this result and lacking detailed evaluation data for three SWE systems under GPT-4o~\cite{openaiHelloGPT4o} and o3-mini~\cite{openaiOpenAIO3mini}, we only show the results of method using DeepSeek-V3.2-Exp, which are shown in Table~\ref{tab:rq1ByRepo}. Our method also demonstrates advantages at the per-project level, achieving the best performance in 12 out of 13 projects, and ranking second in the remaining project, njs. These results demonstrate the strong effectiveness and generalization of our method across diverse real-world projects.

\textbf{Cost.} The average cost per issue for each method is listed in the rightmost column of Table~\ref{tab:rq1Main}. {\tool} demonstrates a relatively low cost, averaging \$0.072 per issue, higher than OpenHands and Aider under the same model. However, these two baselines exhibit significantly poorer performance compared to ours. Specifically, our method can resolve 10 issues for every 1 USD, which is notably higher than all baselines, which can resolve only between 0.098 and 6 issues per dollar. On the other hand, compared to methods built upon other models, {\tool} exhibits a clear cost advantage across all three SWE systems and our method variants that adopt them as base models. This advantage stems from the cost-effective nature of DeepSeek-V3.2-Exp, which employs the DeepSeek Sparse Attention~\cite{githubDeepSeekV32ExpDeepSeek_V3_2pdfMain} mechanism to substantially reduce inference computation overhead, thereby lowering API costs.

\subsection{RQ2: {\shortSecondRQ}}

\newcommand{\modelname}{\text{\textit{\tool}}}
\newcommand{\modelbase}{${\modelname}_\text{\textit{base}}$\xspace}
\newcommand{\modelprecc}{${\modelname}_\text{\textit{cpc}}$\xspace}
\newcommand{\modelspa}{${\modelname}_\text{\textit{spa}}$\xspace}
\newcommand{\modelbasePreccSpa}{${\modelname}_\text{\textit{{base,cpc,spa}}}$\xspace}
\newcommand{\modelfull}{${\modelname}_\text{\textit{full}}$\xspace}
\newcommand{\modelenhvulnloc}{${\modelname}_\text{\textit{enhanceVulnLoc}}$\xspace}
\newcommand{\modelenhpatchgen}{${\modelname}_\text{\textit{enhancePatchGen}}$\xspace}
\newcommand{\modelbasesani}{${\modelname}_\text{\textit{base}}^\text{\textit{sanitizer}}$\xspace}
\newcommand{\modelfullsani}{${\modelname}_\text{\textit{full}}^\text{\textit{sanitizer}}$\xspace}

\newcommand{\modelbasesimplevoting}{${\modelname}_\text{\textit{base}}^\text{\textit{simpleVoting}}$\xspace}
\newcommand{\modelfullsimplevoting}{${\modelname}_\text{\textit{full}}^\text{\textit{simpleVoting}}$\xspace}

\begin{table}

\centering
\caption{Comprehensive ablation study of \tool on \secBenchFirstEighty. Here \%R denotes \%Resolved.}
\label{tab:rq2Main}

\resizebox{\linewidth}{!}{
\begin{tabular}{l|cc|c|l}
\toprule
\multirow{2}{*}{\textbf{Method Variants}} & \multicolumn{2}{c|}{\multirow{2}{*}{\textbf{Design Choices}}} & \multirow{2}{*}{\textbf{\%R}} & \multirow{2}{*}{\textbf{Avg.\$}} \\
& & & & \\
\midrule
\midrule
\cellcolor[HTML]{E7E6E6}\textbf{Two Agents (\S\ref{sec:rq2TwoAgents})} &
  \textbf{CPCAgent} &
  \textbf{SPAAgent} &
  \multicolumn{2}{c}{\cellcolor[HTML]{E7E6E6}} \\
\hline
\modelbase             &   &   & \cellcolor[HTML]{FCFCFF}48.8\% & \$0.0296 \\
\modelprecc            & \ding{51} &   & \cellcolor[HTML]{B4DFC1}61.3\% & \$0.0431 \\
\modelspa              &   & \ding{51} & \cellcolor[HTML]{88CD9B}68.8\% & \$0.0640 \\
\modelfull             & \ding{51} & \ding{51} & \cellcolor[HTML]{63BE7B}75.0\% & \$0.0718 \\
\midrule
\midrule
\cellcolor[HTML]{E7E6E6}\textbf{Stage Enhance.  (\S\ref{sec:rq2StageWiseEnhancement})} &
  \textbf{Vuln. Loc.} &
  \textbf{Patch Gen.} &
  \multicolumn{2}{c}{\cellcolor[HTML]{E7E6E6}} \\
\hline
\modelbase             &   &   & \cellcolor[HTML]{FCFCFF}48.8\% & \$0.0296 \\
\modelenhvulnloc       & \ding{51} &   & \cellcolor[HTML]{E7F4ED}52.5\% & \$0.0723 \\
\modelenhpatchgen      &   & \ding{51} & \cellcolor[HTML]{81CA95}70.0\% & \$0.0725 \\
\modelfull             & \ding{51} & \ding{51} & \cellcolor[HTML]{63BE7B}75.0\% & \$0.0718 \\
\midrule
\midrule
\cellcolor[HTML]{E7E6E6}\textbf{Patch Selection (\S\ref{sec:rq2inputTypes})} &
  \textbf{Simple} &
  \textbf{PoC-based} &
  \multicolumn{2}{c}{\cellcolor[HTML]{E7E6E6}} \\
\hline
\modelfullsimplevoting & \ding{51} &   & \cellcolor[HTML]{FCFCFF}62.5\% & \$0.0718 \\
\modelfull             &   & \ding{51} & \cellcolor[HTML]{63BE7B}75.0\% & \$0.0718 \\
\midrule
\midrule
\cellcolor[HTML]{E7E6E6}\textbf{Input Type  (\S\ref{sec:rq2inputTypes})} &
  \textbf{Sanitizer Log} &
  \textbf{Issue Report} &
  \multicolumn{2}{c}{\cellcolor[HTML]{E7E6E6}} \\
\hline
\modelfullsani         & \ding{51} &   & \cellcolor[HTML]{FCFCFF}63.8\% & \$0.0703 \\
\modelfull             &  \ding{51} & \ding{51} & \cellcolor[HTML]{63BE7B}75.0\% & \$0.0718 \\
\bottomrule
\end{tabular}
} %
\end{table}

In this RQ, we evaluate our design through a comprehensive ablation study, as listed in Table~\ref{tab:rq2Overview}.

\begin{table}[H]
\centering
\caption{A Comprehensive Ablation Study}
\label{tab:rq2Overview}
\begin{tabular}{|c|c|l|}
\hline
\textbf{Analysis} & \textbf{Section} & \textbf{Results} \\
\hline
Agent Contributions & \S\ref{sec:rq2TwoAgents} & Table~\ref{tab:rq2Main} (Top) \\
\hline
Stage-Wise Enhancement & \S\ref{sec:rq2StageWiseEnhancement} & Table~\ref{tab:rq2Main} (Second) \\
\hline
Patch Selection Strategy & \S\ref{sec:rq2ImpactOfPatchSelection} & Table~\ref{tab:rq2Main} (Third) \\
\hline
Input Type Comparison & \S\ref{sec:rq2inputTypes} & Table~\ref{tab:rq2Main} (Bottom) \\
\hline
Patch Space Size & \S\ref{sec:rq2ImpactOfT} & Figure~\ref{fig:rq2ImpactOfT} \\
\hline
Base Model & \S\ref{sec:rq2BaseLLMs} & Table~\ref{tab:rq2LLMs} \\
\hline
Tool Usage & \S\ref{sec:rq2StatsOfToolUsage} & Figure~\ref{fig:rq2StatsOfToolUsage} \\
\hline
\end{tabular}
\end{table}

(1) We evaluate the contributions of \preCCAgent (\textit{\shortPreCCAgent}) and \spaAgent (\textit{\shortSPAAgent}) by enabling or disabling them in various combinations (\S\ref{sec:rq2TwoAgents}, \textit{top section} of Table~\ref{tab:rq2Main}).

(2) Our framework applies enhanced issue reports in two stages: vulnerability localization (\textit{Vuln. Loc.}) and patch generation (\textit{Patch Gen.}). We analyze the stage-wise effect by selectively enabling enhancement at either or both stages (\S\ref{sec:rq2StageWiseEnhancement}, \textit{second section} of Table~\ref{tab:rq2Main}).

(3) We examine the impact of patch selection strategies on performance. Specifically, we design a simple strategy (\textit{Simple}) based on majority voting without PoC validation. We replace the \textit{PoC-based} strategy in {\modelfull} with this variant (\S\ref{sec:rq2ImpactOfPatchSelection}, \textit{third section} of Table~\ref{tab:rq2Main}).

(4) To our knowledge, {\tool} is the first to address vulnerability issues using \textit{issue reports}. Previous methods~\cite{yu2025patchagent, kim2025SAN2PATCH} rely on \textit{sanitizer logs}. We compare the two input types by replacing issue reports with sanitizer logs, highlighting the higher-level semantic and contextual value of issue reports from bug reporters (\S\ref{sec:rq2inputTypes}, \textit{bottom section} of Table~\ref{tab:rq2Main}). 

(5) As described in Section~\ref{sec:impl}, {\tool} generates $T$=5 patches per issue. We conduct experiments with $T$ between 1 and 10, recording the number of issues fixed and average cost as $T$ varies (\S\ref{sec:rq2ImpactOfT}, Figure~\ref{fig:rq2ImpactOfT}).

(6) To assess the impact of the base model, we evaluate variants using different base models (\S\ref{sec:rq2BaseLLMs}, Table~\ref{tab:rq2LLMs}).

(7) We analyze tool-invocation statistics to understand how our agents utilize the tools, following previous LLM agent-based APR work~\cite{Bouzenia2025RepairAgent} (\S\ref{sec:rq2StatsOfToolUsage}, Figure~\ref{fig:rq2StatsOfToolUsage}).

\subsubsection{Effectiveness Analysis of the Two Agents}\label{sec:rq2TwoAgents} To demonstrate the effectiveness of the proposed two agents, we evaluate four variants (top section of Table~\ref{tab:rq2Main}).

\textbf{\modelbase.} Our base framework follows the Agentless workflow~\cite{xia2025agentless}, implementing a two-step issue resolution process. It resolves 48.8\% of issues, outperforming SWE systems from RQ1 (e.g., SWE-agent with the same base model resolves 37.5\%), despite not employing complex agents or dynamic tool calls. This demonstrates that, in the context of vulnerability repair, an LLM-based workflow can achieve strong results even without agentic capabilities and at very low cost (\$0.0296 vs. SWE-agent's \$0.1080), providing a solid foundation for our approach.

\textbf{\modelprecc~vs.~\modelbase.} {\tool} with only {\shortPreCCAgent} successfully resolves 61.3\% of issues, repairing 10 more issues than \modelbase, representing a 25.6\% improvement. This demonstrates the effectiveness of our {\shortPreCCAgent}, which leverages a preliminary context collection phase. By integrating flexible code search and dependency analysis tools to generate a context analysis report, it substantially enhances issue repair performance. Compared with SWE-agent, when our workflow is augmented with similar adaptive code search capabilities, our method significantly outperforms it by 63.3\%.

\textbf{\modelspa~vs.~\modelbase.} {\tool} with only {\shortSPAAgent} resolves 68.8\% of issues, 16 more than \modelbase, an 41.0\% improvement, demonstrating the effectiveness of {\shortSPAAgent}. The agent generates safety property assertions to iteratively understand the vulnerability, identifies violations of security constraints, and thereby achieves a precise comprehension of unsafe behavior, leading to substantial performance gains. Its standalone application outperforms \modelprecc, highlighting the unique and critical contribution of this component.

\textbf{\modelfull~vs.~\modelbasePreccSpa.} When both agents are applied, \modelfull achieves the best performance, successfully fixing 75.0\% of issues, a \textbf{53.8\% improvement} over \modelbase. This demonstrates the effectiveness of our overall design. Moreover, compared with \modelprecc and \modelspa, \modelfull also outperforms them by 22.4\% and 9.1\%, respectively, highlighting the complementary effect of the two agents and confirming that both are essential.

\textbf{Cost Analysis.} As shwon in the \textit{Avg.\$} column of Table~\ref{tab:rq2Main}, \modelbase incurs the lowest cost at \$0.0296 per issue. Adding \preCCAgent increases it to \$0.0431, and \spaAgent alone to \$0.0640. The full system with both agents, \modelfull, reaches \$0.0718 per issue. Despite this increase, the additional cost is justified by the substantial improvement in resolution rate: 75.0\% for \modelfull versus 48.8\% for \modelbase. Notably, even with both agents, the cost remains lower than that of SWE-agent, the fully agent-based and most competitive method, while achieving significantly better performance. The trade-off highlights that both agents effectively improve repair outcomes at moderate computational expense.

\subsubsection{Impact of Stage-Wise Enhancement}\label{sec:rq2StageWiseEnhancement} To examine the effect of applying the enhanced issue report at different stages, we evaluate four variants corresponding to using it in vulnerability localization, patch generation, both, or neither (the second section of Table~\ref{tab:rq2Main}).

\textbf{\modelenhvulnloc.} {\tool} with the enhanced issue report applied solely during the vulnerability localization stage resolves 52.5\% of issues, representing a 7.7\% improvement over \modelbase. This indicates that enhancement during localization yields modest benefits, yet the performance remains 30.0\% lower than that of \modelfull, emphasizing the necessity of also incorporating enhancement in the patch generation stage.

\textbf{\modelenhpatchgen.} 
{\tool} with enhancement applied exclusively during patch generation resolves 70.0\% of issues, an 43.6\% improvement over \modelbase. This highlights the crucial role of enhancement in correct patch generation. However, it still lags 6.7\% behind \modelfull, confirming the complementary benefits of two-stage enhancement.

\textbf{\modelfull.} When the enhanced report is applied to both stages, \modelfull achieves the best performance, underscoring the importance of two-stage enhancement for better issue resolution.

\textbf{Cost Analysis.} We observe that the latter three variants, \modelenhvulnloc, \modelenhpatchgen, and \modelfull, exhibit comparable repair costs (all within \$0.071–\$0.073 per issue). This occurs because the agent-based report generation, which is performed regardless of the stage used, dominates token consumption (70.45\% in \modelfull), while the reports themselves contribute negligibly (2.19\% of all input tokens in \modelfull). Given their performance differences, applying enhancement to both stages maximizes repair effectiveness at a comparable cost, confirming the effectiveness of our design.

\subsubsection{Impact of Patch Selection Strategy}\label{sec:rq2ImpactOfPatchSelection} To evaluate this, we compare two variants that use different patch selection strategies: \textit{direct voting} and \textit{PoC-based voting} (the third section of Table~\ref{tab:rq2Main}).

\textbf{\modelfullsimplevoting~vs.~\modelfull.} Replacing the PoC-based patch selection strategy with the simple voting strategy results in a significant performance drop: \modelfullsimplevoting resolves 62.5\% of issues, 16.7\% lower than \modelfull. This underscores the importance of the PoC-based patch selection strategy, which helps filter out patches that still trigger sanitizers after application. We further analyze the issues with at least one patch selected using the PoC-based voting strategy (63 issues), slightly higher than the number resolved (60 issues). However, three of these issues, although they did not trigger sanitizers, still failed to be resolved due to functionality breaks. This motivates us to improve {\tool} to generate and select more accurate patches in future work.

\begin{figure}
  \centering
  \includegraphics[width=0.9\linewidth]{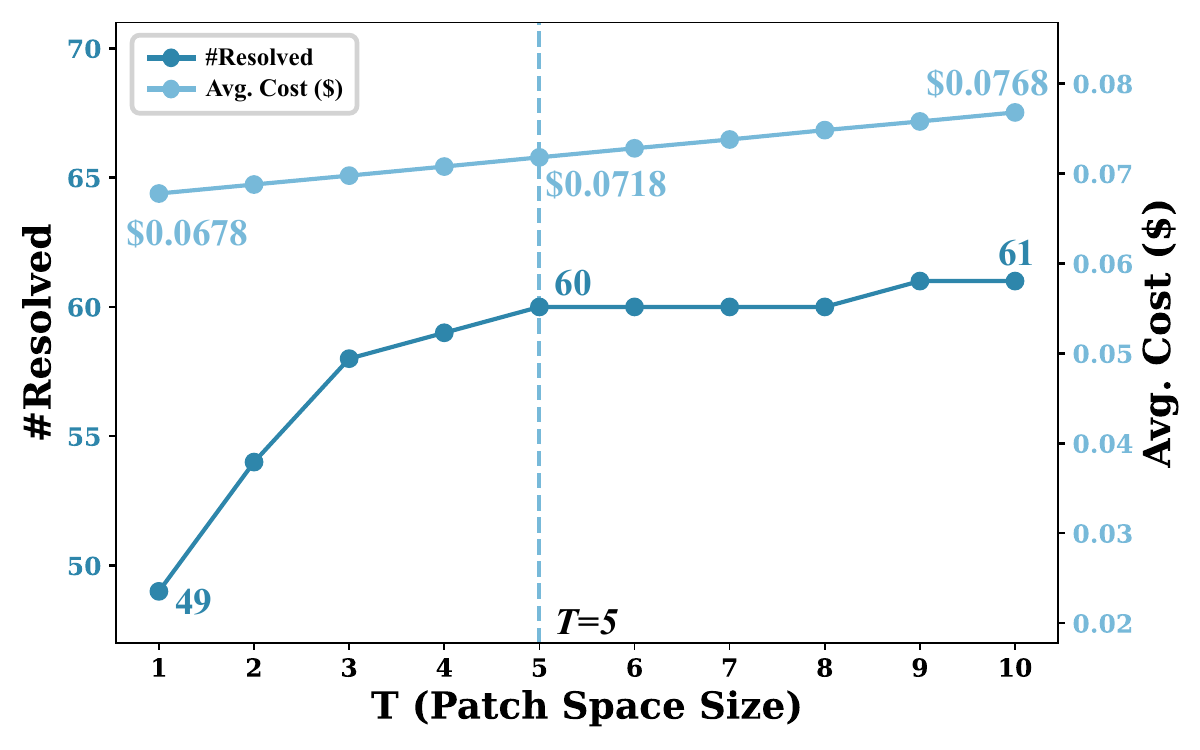}
  \caption{Performance vs. Average Cost for Different Patch Space Sizes}
  \label{fig:rq2ImpactOfT}
\end{figure}

\subsubsection{Impact of Input Types}\label{sec:rq2inputTypes} To assess this, we compare two variants using \textit{sanitizer logs} and \textit{issue reports} (bottom section of Table~\ref{tab:rq2Main}).

\textbf{\modelfullsani~vs.~\modelfull.} Replacing issue reports with sanitizer logs results in a clear performance drop: \modelfullsani fixes 63.8\% of issues, 15.0\% lower than \modelfull, which suggests that issue reports, originating from human reporters, provide richer semantic and contextual information than low-level runtime logs, greatly benefiting the repair process. 
This further underscores the value of studying vulnerability repair based on real-world issue reports as a distinct and valuable research direction.

\subsubsection{Impact of Patch Space Size}\label{sec:rq2ImpactOfT}

To assess the impact of $T$ on performance, we vary $T$ from 1 to 10 and evaluate both the performance and the average cost per issue. The results are shown in Figure~\ref{fig:rq2ImpactOfT}. As $T$ increases from 1 to 5, performance improves from 49 to 60 issues resolved. However, beyond $T$=5, performance plateaus, with only a slight improvement from $T$=8 to $T$=9, and no further gains at $T$=10. In contrast, the cost steadily increases by 7.0\%, from \$0.0718 at $T$=5 to \$0.0768 at $T$=10. Thus, $T$=5 offers a good balance between performance and cost. Notably, even at $T$=1, {\tool} resolves 49 issues, achieving a 61.3\% repair rate, outperforming all RQ1 baselines and further demonstrating the effectiveness of our approach.

\subsubsection{Impact of Base Model}
\label{sec:rq2BaseLLMs} Based on RQ1, we further evaluate {\modelbase} with different base models to determine if our design \textbf{consistently} enhances the base workflow, as shown in Table~\ref{tab:rq2LLMs}.

\begin{table}[H]
\centering
\caption{{\modelbase} vs. {\modelfull} across base models}
\label{tab:rq2LLMs}
\begin{tabular}{l|cc|cc}
\toprule
\multicolumn{1}{c|}{\multirow{2}{*}{\textbf{Model}}} & \multicolumn{4}{c}{\textbf{\%Resolved \& Avg.\$}}                   \\
\multicolumn{1}{c|}{} & \multicolumn{2}{c|}{\textbf{{\modelbase}}} & \multicolumn{2}{c}{\textbf{{\modelfull}}} \\
\midrule
OpenAI GPT-4o                                       & 33.8\% & \$0.30 & 51.3\%\increase{51.9\%} & \$0.85 \\
OpenAI o3-mini                                      & 35.0\% & \$0.17 & 57.5\%\increase{64.3\%} & \$0.33 \\
DeepSeek-V3.2-Exp                                   & 48.8\% & \$0.03 & 75.0\%\increase{53.8\%} & \$0.07 \\
\bottomrule
\end{tabular}
\end{table}

We observe that DeepSeek-V3.2-Exp achieves the highest performance across all models. More importantly, \modelfull consistently outperforms \modelbase, with resolution rates improving by 51.9\%, 64.3\%, and 53.8\% for GPT-4o, o3-mini, and DeepSeek-V3.2-Exp, respectively. This further demonstrates the effectiveness and generalizability of our approach.

\subsubsection{Statistical Analysis of Tool Usage}\label{sec:rq2StatsOfToolUsage}

\begin{figure}
  \centering
  \includegraphics[width=0.9\linewidth]{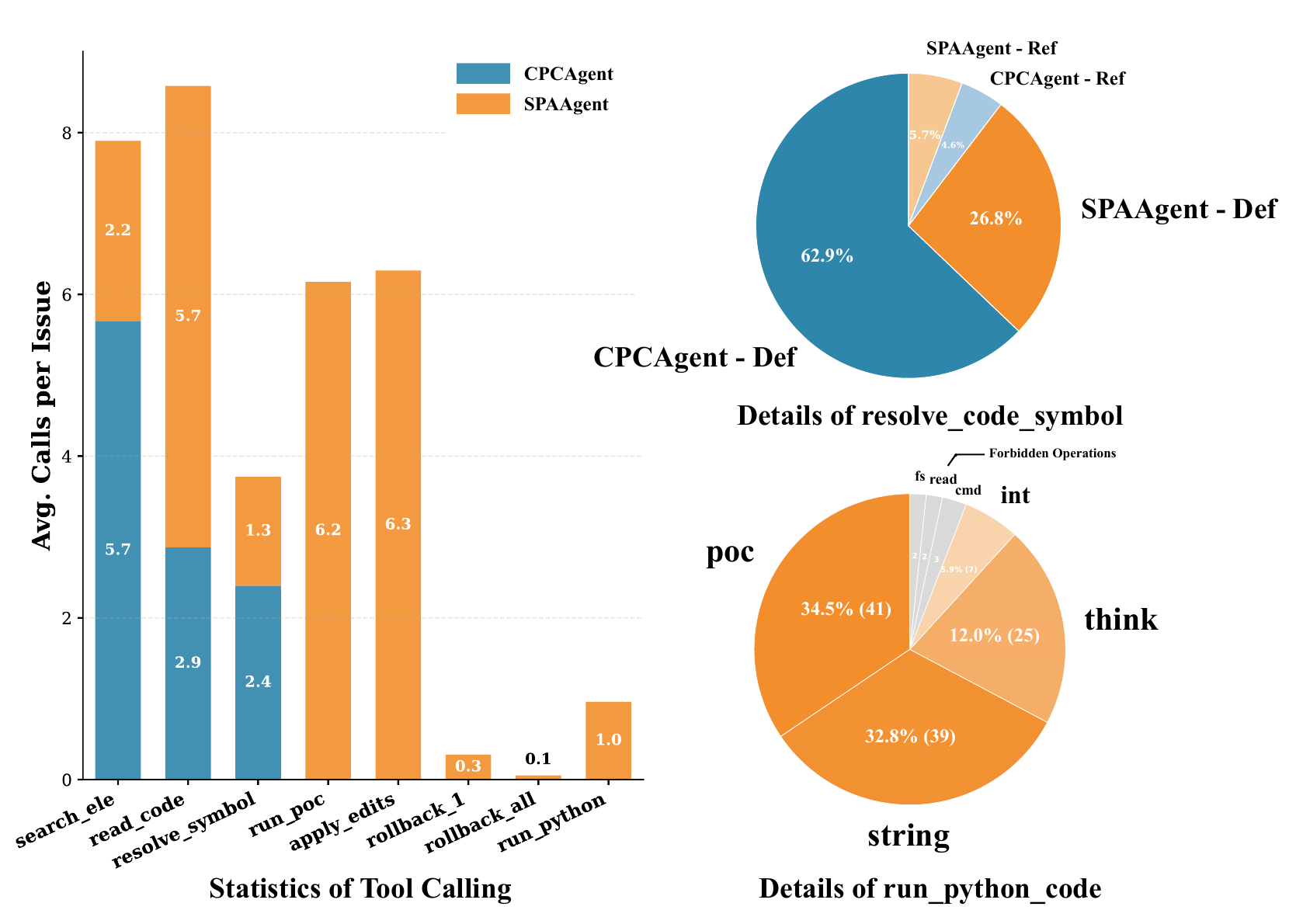}
  \caption{Tool Usage Statistics of {\modelfull}}
  \label{fig:rq2StatsOfToolUsage}
\end{figure}

We analyze how our agents employ toolkits by measuring the average tool invocations per issue, including tool frequencies for {\shortPreCCAgent} and {\shortSPAAgent}. Results in Figure~\ref{fig:rq2StatsOfToolUsage} show an average of 34 calls per issue, covering code retrieval, analysis, and safety property generation, applying and validation.

As shown in the bar chart, {\shortPreCCAgent} uses {\SearchCodeElementInFile}, {\ReadCodeInFile}, and {\ResolveCodeSymbol} for \textit{static} analysis, while {\shortSPAAgent} frequently invokes {\RunPoC}, {\ApplyEdits}, and {\RunPythonCode} for \textit{dynamic} property analysis. On average, {\shortPreCCAgent} performs 5.7 searches, 2.9 reads, and 2.4 symbol resolutions per issue; {\shortSPAAgent} executes PoC 6.2 times, applies 6.3 edits, and runs Python 0.96 times.

\textbf{Code Symbol Resolution.} The upper-right pie chart in Figure~\ref{fig:rq2StatsOfToolUsage} shows that {\shortPreCCAgent} performs 62.9\% definition (\textit{\shortPreCCAgent-Def}) and 5.7\% reference (\textit{\shortPreCCAgent-Ref}) resolutions, while {\shortSPAAgent} allocates 26.8\% and 4.6\% to the same categories, respectively.

\textbf{Python Code Execution.} The lower-right pie chart summarizes \texttt{\RunPythonCode} invocations, each manually labeled by functionality. Among them, 34.5\% retrieve \textit{poc} outputs (e.g., calling \texttt{get\_poc\_output()}), 32.8\% perform \textit{string} operations (e.g., parsing string with regex). Interestingly, 12.0\% exhibit a \textit{think} tag, where generated Python code prints hardcoded text without computation. In fact, these calls, though non-functional, externalize the agent's intermediate reasoning, similar to the ``think'' mechanisms in OpenHands~\cite{wang2024openhands}, Anthropic's recent practice~\cite{anthropic2025thinkTool}, and ByteDance's TRAE Agent~\cite{bytedance2025TRAE} that facilitate complex multi-tool coordination. Another 5.9\% involve integer arithmetic (e.g., range comparisons, arithmetic, or bitwise operations), and seven attempts included forbidden actions such as file system access (\textit{fs}), file reading (\textit{read}), or command execution (\textit{cmd}), all correctly blocked by the sandboxed execution environment.

Overall, {\tool} exhibits phase-dependent toolkit usage: {\shortPreCCAgent} focuses on static context collection, while {\shortSPAAgent} emphasizes property analysis, generation, and dynamic validation.

\subsection{RQ3: {\shortThirdRQ}}

\begin{table}
\centering
\caption{Issue resolution performance on \secBenchFull.}
\label{tab:rq3Main}
\resizebox{\linewidth}{!}{
\begin{tabular}{rc|cc|ccc}
\toprule
 &
   &
  \multicolumn{2}{c}{\textbf{DeepSeek-V3.2-Exp}} &
  \multicolumn{3}{c}{\textbf{Claude 3.7 Sonnet}} \\
\multirow{-2}{*}{\textbf{Repository}} &
  \multirow{-2}{*}{\textbf{\#Issue}} &
  \textbf{{\tool}} &
  \textbf{Basic Workflow} &
  \textbf{SWE-agent} &
  \textbf{OpenHands} &
  \textbf{Aider} \\
\midrule
exiv2 &
  10 &
  \cellcolor[HTML]{8ACE9C}4(40.0\%) &
  \cellcolor[HTML]{FCFCFF}1(10.0\%) &
  \cellcolor[HTML]{FCFCFF}1(10.0\%) &
  \cellcolor[HTML]{FCFCFF}1(10.0\%) &
  \cellcolor[HTML]{63BE7B}5(50.0\%) \\
faad2 &
  12 &
  \cellcolor[HTML]{63BE7B}8(66.7\%) &
  \cellcolor[HTML]{C3E5CE}3(25.0\%) &
  \cellcolor[HTML]{B0DDBD}4(33.3\%) &
  \cellcolor[HTML]{D6EDDE}2(16.7\%) &
  \cellcolor[HTML]{FCFCFF}0(0.0\%) \\
gpac &
  43 &
  \cellcolor[HTML]{63BE7B}32(74.4\%) &
  \cellcolor[HTML]{8ACE9C}24(55.8\%) &
  \cellcolor[HTML]{B0DDBD}16(37.2\%) &
  \cellcolor[HTML]{FCFCFF}0(0.0\%) &
  \cellcolor[HTML]{DBEFE3}7(16.3\%) \\
imagemagick &
  31 &
  \cellcolor[HTML]{63BE7B}24(77.4\%) &
  \cellcolor[HTML]{9DD6AD}18(58.1\%) &
  \cellcolor[HTML]{E9F5EF}10(32.3\%) &
  \cellcolor[HTML]{77C68C}22(71.0\%) &
  \cellcolor[HTML]{FCFCFF}8(25.8\%) \\
jq &
  1 &
  \cellcolor[HTML]{63BE7B}1(100.0\%) &
  \cellcolor[HTML]{63BE7B}1(100.0\%) &
  \cellcolor[HTML]{63BE7B}1(100.0\%) &
  \cellcolor[HTML]{FCFCFF}0(0.0\%) &
  \cellcolor[HTML]{FCFCFF}0(0.0\%) \\
libarchive &
  3 &
  \cellcolor[HTML]{63BE7B}3(100.0\%) &
  \cellcolor[HTML]{96D3A7}2(66.7\%) &
  \cellcolor[HTML]{FCFCFF}0(0.0\%) &
  \cellcolor[HTML]{FCFCFF}0(0.0\%) &
  \cellcolor[HTML]{96D3A7}2(66.7\%) \\
libdwarf-code &
  2 &
  \cellcolor[HTML]{63BE7B}2(100.0\%) &
  \cellcolor[HTML]{FCFCFF}1(50.0\%) &
  \cellcolor[HTML]{FCFCFF}1(50.0\%) &
  \cellcolor[HTML]{FCFCFF}1(50.0\%) &
  \cellcolor[HTML]{FCFCFF}1(50.0\%) \\
libheif &
  2 &
  \cellcolor[HTML]{FCFCFF}0(0.0\%) &
  \cellcolor[HTML]{63BE7B}1(50.0\%) &
  \cellcolor[HTML]{FCFCFF}0(0.0\%) &
  \cellcolor[HTML]{63BE7B}1(50.0\%) &
  \cellcolor[HTML]{FCFCFF}0(0.0\%) \\
libiec61850 &
  2 &
  \cellcolor[HTML]{63BE7B}1(50.0\%) &
  \cellcolor[HTML]{63BE7B}1(50.0\%) &
  \cellcolor[HTML]{63BE7B}1(50.0\%) &
  \cellcolor[HTML]{63BE7B}1(50.0\%) &
  \cellcolor[HTML]{63BE7B}1(50.0\%) \\
libjpeg-turbo &
  1 &
  \cellcolor[HTML]{FCFCFF}0(0.0\%) &
  \cellcolor[HTML]{FCFCFF}0(0.0\%) &
  \cellcolor[HTML]{FCFCFF}0(0.0\%) &
  \cellcolor[HTML]{63BE7B}1(100.0\%) &
  \cellcolor[HTML]{FCFCFF}0(0.0\%) \\
liblouis &
  1 &
  \cellcolor[HTML]{63BE7B}1(100.0\%) &
  \cellcolor[HTML]{63BE7B}1(100.0\%) &
  \cellcolor[HTML]{63BE7B}1(100.0\%) &
  \cellcolor[HTML]{FCFCFF}0(0.0\%) &
  \cellcolor[HTML]{63BE7B}1(100.0\%) \\
libmodbus &
  1 &
  \cellcolor[HTML]{63BE7B}1(100.0\%) &
  \cellcolor[HTML]{63BE7B}1(100.0\%) &
  \cellcolor[HTML]{FCFCFF}0(0.0\%) &
  \cellcolor[HTML]{63BE7B}1(100.0\%) &
  \cellcolor[HTML]{FCFCFF}0(0.0\%) \\
libplist &
  1 &
  \cellcolor[HTML]{63BE7B}1(100.0\%) &
  \cellcolor[HTML]{63BE7B}1(100.0\%) &
  \cellcolor[HTML]{63BE7B}1(100.0\%) &
  \cellcolor[HTML]{63BE7B}1(100.0\%) &
  \cellcolor[HTML]{63BE7B}1(100.0\%) \\
libredwg &
  20 &
  \cellcolor[HTML]{63BE7B}12(60.0\%) &
  \cellcolor[HTML]{63BE7B}12(60.0\%) &
  \cellcolor[HTML]{E9F5EF}5(25.0\%) &
  \cellcolor[HTML]{9DD6AD}9(45.0\%) &
  \cellcolor[HTML]{FCFCFF}4(20.0\%) \\
libsndfile &
  1 &
  \cellcolor[HTML]{63BE7B}1(100.0\%) &
  \cellcolor[HTML]{63BE7B}1(100.0\%) &
  \cellcolor[HTML]{63BE7B}1(100.0\%) &
  \cellcolor[HTML]{FCFCFF}0(0.0\%) &
  \cellcolor[HTML]{FCFCFF}0(0.0\%) \\
libxls &
  1 &
  \cellcolor[HTML]{63BE7B}1(100.0\%) &
  \cellcolor[HTML]{63BE7B}1(100.0\%) &
  \cellcolor[HTML]{FCFCFF}0(0.0\%) &
  \cellcolor[HTML]{FCFCFF}0(0.0\%) &
  \cellcolor[HTML]{FCFCFF}0(0.0\%) \\
matio &
  7 &
  \cellcolor[HTML]{63BE7B}4(57.1\%) &
  \cellcolor[HTML]{D6EDDE}1(14.3\%) &
  \cellcolor[HTML]{FCFCFF}0(0.0\%) &
  \cellcolor[HTML]{63BE7B}4(57.1\%) &
  \cellcolor[HTML]{D6EDDE}1(14.3\%) \\
md4c &
  3 &
  \cellcolor[HTML]{63BE7B}1(33.3\%) &
  \cellcolor[HTML]{63BE7B}1(33.3\%) &
  \cellcolor[HTML]{FCFCFF}0(0.0\%) &
  \cellcolor[HTML]{FCFCFF}0(0.0\%) &
  \cellcolor[HTML]{FCFCFF}0(0.0\%) \\
mruby &
  21 &
  \cellcolor[HTML]{63BE7B}16(76.2\%) &
  \cellcolor[HTML]{A1D7B0}12(57.1\%) &
  \cellcolor[HTML]{91D1A3}13(61.9\%) &
  \cellcolor[HTML]{B0DDBD}11(52.4\%) &
  \cellcolor[HTML]{FCFCFF}6(28.6\%) \\
njs &
  17 &
  \cellcolor[HTML]{7DC991}9(52.9\%) &
  \cellcolor[HTML]{B0DDBD}7(41.2\%) &
  \cellcolor[HTML]{FCFCFF}4(23.5\%) &
  \cellcolor[HTML]{63BE7B}10(58.8\%) &
  \cellcolor[HTML]{FCFCFF}4(23.5\%) \\
openexr &
  3 &
  \cellcolor[HTML]{C9E8D3}1(33.3\%) &
  \cellcolor[HTML]{63BE7B}3(100.0\%) &
  \cellcolor[HTML]{FCFCFF}0(0.0\%) &
  \cellcolor[HTML]{FCFCFF}0(0.0\%) &
  \cellcolor[HTML]{96D3A7}2(66.7\%) \\
openjpeg &
  5 &
  \cellcolor[HTML]{63BE7B}5(100.0\%) &
  \cellcolor[HTML]{B0DDBD}3(60.0\%) &
  \cellcolor[HTML]{FCFCFF}1(20.0\%) &
  \cellcolor[HTML]{FCFCFF}1(20.0\%) &
  \cellcolor[HTML]{B0DDBD}3(60.0\%) \\
php-src &
  2 &
  \cellcolor[HTML]{63BE7B}2(100.0\%) &
  \cellcolor[HTML]{63BE7B}2(100.0\%) &
  \cellcolor[HTML]{B0DDBD}1(50.0\%) &
  \cellcolor[HTML]{FCFCFF}0(0.0\%) &
  \cellcolor[HTML]{FCFCFF}0(0.0\%) \\
qpdf &
  1 &
  0(0.0\%) &
  0(0.0\%) &
  0(0.0\%) &
  0(0.0\%) &
  0(0.0\%) \\
readstat &
  1 &
  \cellcolor[HTML]{63BE7B}1(100.0\%) &
  \cellcolor[HTML]{FCFCFF}0(0.0\%) &
  \cellcolor[HTML]{FCFCFF}0(0.0\%) &
  \cellcolor[HTML]{FCFCFF}0(0.0\%) &
  \cellcolor[HTML]{FCFCFF}0(0.0\%) \\
upx &
  3 &
  \cellcolor[HTML]{63BE7B}1(33.3\%) &
  \cellcolor[HTML]{FCFCFF}0(0.0\%) &
  \cellcolor[HTML]{63BE7B}1(33.3\%) &
  \cellcolor[HTML]{63BE7B}1(33.3\%) &
  \cellcolor[HTML]{FCFCFF}0(0.0\%) \\
wabt &
  1 &
  \cellcolor[HTML]{63BE7B}1(100.0\%) &
  \cellcolor[HTML]{FCFCFF}0(0.0\%) &
  \cellcolor[HTML]{63BE7B}1(100.0\%) &
  \cellcolor[HTML]{63BE7B}1(100.0\%) &
  \cellcolor[HTML]{63BE7B}1(100.0\%) \\
yaml-cpp &
  1 &
  \cellcolor[HTML]{63BE7B}1(100.0\%) &
  \cellcolor[HTML]{FCFCFF}0(0.0\%) &
  \cellcolor[HTML]{FCFCFF}0(0.0\%) &
  \cellcolor[HTML]{FCFCFF}0(0.0\%) &
  \cellcolor[HTML]{FCFCFF}0(0.0\%) \\
yara &
  3 &
  \cellcolor[HTML]{63BE7B}1(33.3\%) &
  \cellcolor[HTML]{FCFCFF}0(0.0\%) &
  \cellcolor[HTML]{FCFCFF}0(0.0\%) &
  \cellcolor[HTML]{FCFCFF}0(0.0\%) &
  \cellcolor[HTML]{FCFCFF}0(0.0\%) \\
\midrule
\multicolumn{2}{c|}{\textbf{\#Resolved}} &
  \cellcolor[HTML]{63BE7B}135 &
  \cellcolor[HTML]{A4D9B3}98 &
  \cellcolor[HTML]{E1F1E7}63 &
  \cellcolor[HTML]{D8EEE0}68 &
  \cellcolor[HTML]{FCFCFF}47 \\
\multicolumn{2}{c|}{\textbf{\%Resolved}} &
  \cellcolor[HTML]{63BE7B}67.5\% &
  \cellcolor[HTML]{A4D9B3}49.0\% &
  \cellcolor[HTML]{E1F1E7}31.5\% &
  \cellcolor[HTML]{D8EEE0}34.0\% &
  \cellcolor[HTML]{FCFCFF}23.5\% \\
\bottomrule
\end{tabular}
} %

\end{table}

In this RQ, we further conduct evaluation experiments on the {\secBenchFull} by testing our method on the 200 issues. Moreover, to investigate the additional gains brought by our design on the basic workflow, we also evaluate {\modelbase}, as described in RQ2. The overall and per-project issue-fixing results are presented in Table~\ref{tab:rq3Main}.

\textbf{Performance.} {\tool} fixes 135 issues, achieving a solution rate of 67.5\%, outperforming the best-performing SWE system on the SEC-bench leaderboard~\cite{secbenchSECbenchLeaderboard} as of November 2025 (the time of paper submission), OpenHands~\cite{wang2024openhands} with Claude 3.7 Sonnet~\cite{anthropicClaude37Sonnet} (34.0\%) by 98.5\%. More importantly, our agent-enhanced workflow significantly outperforms the {\modelbase}, achieving an 37.8\% improvement. This demonstrates that our approach is equally effective on {\secBenchFull}, further validating its generalizability. Additionally, our method maintains an advantage across different projects, similar to its performance on {\secBenchFirstEighty}, achieving the best results on the majority of projects (23 out of 29). Notably, we observe that the resolve rate on {\secBenchFull} is lower than on {\secBenchFirstEighty}, likely due to the increased difficulty of issues in the full set. As shown in Table~\ref{tab:agentToolkits}, compared to the subset, the full set of issues requires more modifications on average: more files to be modified (1.28 vs. 1.10), more hunks to be altered (2.43 vs. 1.59), and more lines to be edited (17.29 vs. 11.03). This motivates us to adapt our approach for more complex vulnerability repair scenarios in future work.

\textbf{Cost.} Since the SEC-bench paper~\cite{lee2025secBench} does not report the costs of these three baselines over the full dataset, disabling a direct comparison. We still report the costs for the two methods we evaluated. On average, \tool incurs a cost of \$0.0734 per issue, while \modelbase incurs \$0.0274, comparable to the analysis in RQ2.

\section{Threats to Validity}

\textbf{External Validity.} Language Extensibility: Our method is implemented and evaluated only on C/C++. Nevertheless, it can be extended to other programming languages by adapting our agent toolkits, particularly the code element search and code symbol resolution tools, which can be implemented by leveraging language-specific static analysis tools and LSP libraries (e.g, Spoon~\cite{Renaud2015SpoonLibrary} and Eclipse JDT Language Server~\cite{githubGitHubEclipsejdtlseclipsejdtls} for Java).

\textbf{Internal Validity.} (1) Data Leakage: There is a potential risk that developer patches in the benchmark datasets may overlap with the LLM's training data. To mitigate this, we follow the practices established in existing APR studies~\cite{xia2025agentless, huang2024ntr, huang2025guirepair} by evaluating {\tool} using the same base models as the baselines, ensuring that potential data leakage does not unfairly advantage {\tool}. Additionally, we observe significant improvements in performance over the Agentless-based workflow alone, which further mitigates this concern. (2) Hyperparameter Selection: As described in Section~\ref{sec:impl}, we select most hyperparameters (e.g., $N$, $M$, temperature, chunk size, text overlap) based on the experience from Agentless~\cite{xia2025agentless}. The patch space size, $T$, is empirically set to 5 based on preliminary experiments and practical considerations. And the impact of this choice is further explored in RQ2 (\S\ref{sec:rq2ImpactOfT}). (3) Security Threats: To mitigate security risks, we isolate vulnerability PoCs and agent-generated Python code execution in separate Docker containers, respectively, preventing potential threats from affecting the host system. (4) API Cost: The use of LLMs may incur higher API costs. To address this, we evaluate the average cost per issue and demonstrate that the performance gains justify the additional overhead. Moreover, our approach maintains a cost-effective advantage compared to existing LLM-based SWE baselines.

\section{Related Work}

\subsection{Software Issue Resolution}

Automated Program Repair (APR), which aims to automatically fix software defects~\cite{liu2019tbar, le2012GenProg}, includes issue resolution as a subset. Recent advances in LLM-based APR methods have shown promising results~\cite{xia2024chatRepair, yin2024thinkrepair, bouzenia2024repairagent, zhang2025ReinFix}. SWE-bench~\cite{jimenez2023swebench}, a recently proposed benchmark for evaluating LLM-based systems on GitHub issues in real-world repositories, has further advanced practical APR techniques~\cite{yang2024sweAgent, wang2024openhands, gauthier2024aider, bytedance2025TRAE, liu2024marscode, xia2025agentless, ruan2025SpecRover, tao2024magis, zhang2024autocoderover, meng2024empiricalStudyOnLLMBasedBugFixing, chenYeHe2025prometheus}, which can be classified as agent-based or workflow-based. Agent-based methods, like SWEAgent~\cite{yang2024sweAgent} and OpenHands~\cite{wang2024openhands}, use tools like command and code execution to enable LLM agents to interact with the environment. In contrast, workflow-based methods, such as Agentless~\cite{xia2025agentless}, resolve issues through steps like fault localization and patch generation, without relying on agent-based tools, yet still achieve competitive performance. Our approach combines these paradigms, creating a hybrid agent system that balances flexibility and determinism. Beyond common issue resolution, specialized efforts such as SWE-bench Multimodal~\cite{yang2024sweBenchMultimodal}, which evaluates systems on resolving issues in visual JavaScript applications, have emerged. One such effort, GUIRepair~\cite{huang2025guirepair}, uses multimodal LLMs with an agentless framework to address these issues. Another example is AGENTISSUE-BENCH~\cite{rahardja2025canAgentsFixAgentIssues}, which evaluates agent systems in resolving issues within other agents. Our method focuses on vulnerability issue resolution, using the vulnerability-specific method to fix such issues. We evaluate it on SEC-bench~\cite{lee2025secBench}, the first vulnerability issue resolution benchmark, achieving state-of-the-art performance.

\subsection{Automated Vulnerability Repair}

Automated Vulnerability Repair (AVR), a subset of APR, focuses on fixing vulnerabilities rather than common bugs~\cite{li2025sokAVR}. Several LLM-based AVR methods~\cite{fu2022vulrepair, zhou2024vulMaster, pearce2023examiningZeroShotVulRepairWithLLMs, chen2023vrepair, nong2025appatch} have shown promise but often still rely on human-provided fault locations or CWE labels (e.g., VRepair~\cite{chen2023vrepair}, VulRepair~\cite{fu2022vulrepair}, APPATCH~\cite{nong2025appatch}). Methods such as PatchAgent~\cite{yu2025patchagent} and SAN2PATCH~\cite{kim2025SAN2PATCH} automate fault localization and patch generation using sanitizer logs but overlook the semantic context in issue reports, which is critical for understanding vulnerabilities, as shown in our RQ2 (\S\ref{sec:rq2inputTypes}). Furthermore, these approaches lack AVR-specific designs and are not substantially different from general APR systems. For example, PatchAgent's use of tools for context retrieval in vulnerability repositories is similar to the functions of general-purpose agents like SWE-agent~\cite{yang2024sweAgent} and MarsCoder~\cite{liu2024marscode}, which also employ tools like LSP for adaptive repository exploration. This lack of domain-specific design raises questions about whether these methods~\cite{chen2023vrepair, fu2022vulrepair, yu2025patchagent, kim2025SAN2PATCH} provide any advantages over existing general state-of-the-art APR systems~\cite{yang2024sweAgent, wang2024openhands, bytedance2025TRAE, xia2024chatRepair, zhang2025ReinFix}.

In contrast, our approach focuses on a key characteristic of vulnerabilities: they often arise due to violations of safety properties~\cite{useSafePropToGenVulPatch}. We propose {\shortSPAAgent}, which uses LLMs to iteratively understand target vulnerabilities and generate properties. By mining and validating properties, SPAAgent provides deeper insights into vulnerabilities, enhancing fault localization and repair. This design significantly outperforms {\modelprecc} without property-based analysis, which resembles PatchAgent and SWE-agent that rely on general context-retrieval tools for vulnerability repair (RQ2, \S\ref{sec:rq2TwoAgents}). Consequently, our approach not only automates vulnerability issue resolution but also improves performance through a deeper exploration of vulnerability-specific characteristics.

Before the era of LLMs, AVR techniques like Senx~\cite{useSafePropToGenVulPatch} used symbolic execution to mine and apply safety properties. However, (1) Senx only supports three predefined bug types (buffer overflows, bad casts, and integer overflows) with corresponding manually crafted property templates, limiting its applicability. In contrast, our approach leverages LLMs to autonomously discover properties for the target vulnerability, avoiding predefined constraints. (2) Senx's symbolic execution lacks semantic understanding of source code and does not handle unstructured bug reports well. (3) Symbolic execution suffers from path explosion and performance issues. In our initial experiments, we observed that symbolic execution took far longer than running a PoC, reducing its practicality. This led us to abandon symbolic execution in favor of directly leveraging LLM-based agents to edit repositories and dynamically execute PoCs, enabling iterative feedback for property generation.

In summary, our approach combines general agent-based exploration methods for common bugs with vulnerability-specific agent-based property analysis, resulting in an effective and vulnerability-specific program repair method.

\section{Conclusion}

In this work, we propose {\tool}, a novel approach for automated vulnerability issue resolution that combines the flexibility of agent-based systems with the stability of workflow-driven repair. Specifically, {\tool} integrates two specialized agents for adaptive context pre-collection and safety property-based reasoning, enabling more effective vulnerability localization and patch generation. Our experiments show that {\tool} significantly outperforms existing methods, highlighting the effectiveness of our approach in real-world vulnerability repair.

\bibliographystyle{IEEEtran}
\bibliography{reference}

@inproceedings{xia2024chatRepair,
  title={Automated program repair via conversation: Fixing 162 out of 337 bugs for \$0.42 each using ChatGPT},
  author={Xia, Chunqiu Steven and Zhang, Lingming},
  booktitle={Proceedings of the 33rd ACM SIGSOFT International Symposium on Software Testing and Analysis},
  pages={819--831},
  year={2024}
}

@article{liu2024marscode,
  title={Marscode agent: Ai-native automated bug fixing},
  author={Liu, Yizhou and Gao, Pengfei and Wang, Xinchen and Liu, Jie and Shi, Yexuan and Zhang, Zhao and Peng, Chao},
  journal={arXiv preprint arXiv:2409.00899},
  year={2024}
}

@inproceedings{yu2025patchagent,
  title={PatchAgent: A practical program repair agent mimicking human expertise},
  author={Yu, Zheng and Guo, Ziyi and Wu, Yuhang and Yu, Jiahao and Xu, Meng and Mu, Dongliang and Chen, Yan and Xing, Xinyu},
  booktitle={Proceedings of the 34th USENIX Security Symposium (USENIX Security’25), Seattle, WA, USA},
  year={2025}
}

@inproceedings{wang2024codeact,
  title={Executable code actions elicit better llm agents},
  author={Wang, Xingyao and Chen, Yangyi and Yuan, Lifan and Zhang, Yizhe and Li, Yunzhu and Peng, Hao and Ji, Heng},
  booktitle={Forty-first International Conference on Machine Learning},
  year={2024}
}

@article{huang2025guirepair,
  title={Seeing is Fixing: Cross-Modal Reasoning with Multimodal LLMs for Visual Software Issue Fixing},
  author={Huang, Kai and Zhang, Jian and Xie, Xiaofei and Chen, Chunyang},
  journal={arXiv preprint arXiv:2506.16136},
  year={2025}
}

@article{xia2025agentless,
  title={Demystifying llm-based software engineering agents},
  author={Xia, Chunqiu Steven and Deng, Yinlin and Dunn, Soren and Zhang, Lingming},
  journal={Proceedings of the ACM on Software Engineering},
  volume={2},
  number={FSE},
  pages={801--824},
  year={2025},
  publisher={ACM New York, NY, USA}
}

@misc{gauthier2024aider,
  title={Aider is ai pair programming in your terminal},
  author={Gauthier, Paul and Contributors, Aider-AI},
  year={2024},
  url={https://aider.chat}
}

@article{lee2025secBench,
  title={SEC-bench: Automated Benchmarking of LLM Agents on Real-World Software Security Tasks},
  author={Lee, Hwiwon and Zhang, Ziqi and Lu, Hanxiao and Zhang, Lingming},
  journal={arXiv preprint arXiv:2506.11791},
  year={2025}
}

@article{bytedance2025TRAE,
  title={Trae agent: An llm-based agent for software engineering with test-time scaling},
  author={Gao, Pengfei and Tian, Zhao and Meng, Xiangxin and Wang, Xinchen and Hu, Ruida and Xiao, Yuanan and Liu, Yizhou and Zhang, Zhao and Chen, Junjie and Gao, Cuiyun and others},
  journal={arXiv preprint arXiv:2507.23370},
  year={2025}
}

@inproceedings{ruan2025SpecRover,
author = {Ruan, Haifeng and Zhang, Yuntong and Roychoudhury, Abhik},
title = {SpecRover: Code Intent Extraction via LLMs},
year = {2025},
isbn = {9798331505691},
publisher = {IEEE Press},
url = {https://doi.org/10.1109/ICSE55347.2025.00080},
doi = {10.1109/ICSE55347.2025.00080},
booktitle = {Proceedings of the IEEE/ACM 47th International Conference on Software Engineering},
pages = {963–974},
numpages = {12},
location = {Ottawa, Ontario, Canada},
series = {ICSE '25}
}

@article{yang2024sweAgent,
  title={Swe-agent: Agent-computer interfaces enable automated software engineering},
  author={Yang, John and Jimenez, Carlos E and Wettig, Alexander and Lieret, Kilian and Yao, Shunyu and Narasimhan, Karthik and Press, Ofir},
  journal={Advances in Neural Information Processing Systems},
  volume={37},
  pages={50528--50652},
  year={2024}
}

@misc{anthropicClaude37Sonnet,
	author = {Anthropic},
	title = {{C}laude 3.7 {S}onnet and {C}laude {C}ode},
	howpublished = {\url{https://www.anthropic.com/news/claude-3-7-sonnet}},
	year = {2025},
}

@misc{openaiHelloGPT4o,
	author = {OpenAI},
	title = {{H}ello {G}{P}{T}-4o},
	howpublished = {\url{https://openai.com/index/hello-gpt-4o}},
	year = {2024},
}

@misc{openaiOpenAIO3mini,
	author = {OpenAI},
	title = {{O}pen{A}{I} o3-mini},
	howpublished = {\url{https://openai.com/index/openai-o3-mini/}},
	year = {2025},
}

@misc{deepseekIntroducingDeepSeekV32Exp,
	author = {DeepSeek},
	title = {{I}ntroducing {D}eep{S}eek-{V}3.2-{E}xp},
	howpublished = {\url{https://api-docs.deepseek.com/news/news250929}},
	year = {2025},
}

@article{wang2024openhands,
  title={Openhands: An open platform for ai software developers as generalist agents},
  author={Wang, Xingyao and Li, Boxuan and Song, Yufan and Xu, Frank F and Tang, Xiangru and Zhuge, Mingchen and Pan, Jiayi and Song, Yueqi and Li, Bowen and Singh, Jaskirat and others},
  journal={arXiv preprint arXiv:2407.16741},
  year={2024}
}

@inproceedings{kim2025SAN2PATCH,
  title={Logs In, Patches Out: Automated Vulnerability Repair via $\{$Tree-of-Thought$\}$$\{$LLM$\}$ Analysis},
  author={Kim, Youngjoon and Shin, Sunguk and Kim, Hyoungshick and Yoon, Jiwon},
  booktitle={34th USENIX Security Symposium (USENIX Security 25)},
  pages={4401--4419},
  year={2025}
}

@inproceedings{
yao2023react,
title={ReAct: Synergizing Reasoning and Acting in Language Models},
author={Shunyu Yao and Jeffrey Zhao and Dian Yu and Nan Du and Izhak Shafran and Karthik R Narasimhan and Yuan Cao},
booktitle={The Eleventh International Conference on Learning Representations },
year={2023},
url={https://openreview.net/forum?id=WE_vluYUL-X}
}

@INPROCEEDINGS{Bouzenia2025RepairAgent,
  author={Bouzenia, Islem and Devanbu, Premkumar and Pradel, Michael},
  booktitle={2025 IEEE/ACM 47th International Conference on Software Engineering (ICSE)}, 
  title={RepairAgent: An Autonomous, LLM-Based Agent for Program Repair}, 
  year={2025},
  volume={},
  number={},
  pages={2188-2200},
  doi={10.1109/ICSE55347.2025.00157}}

@misc{anthropic2025thinkTool,
	author = {Anthropic},
	title = {{T}he "think" tool: {E}nabling {C}laude to stop and think},
	howpublished = {\url{https://www.anthropic.com/engineering/claude-think-tool}},
	year = {2025},
}

@misc{gitscmDiffformatDocumentation,
	author = {Git Community},
	title = {{G}it - diff-format {D}ocumentation},
	howpublished = {\url{https://git-scm.com/docs/diff-format}},
	year = {},
	note = {[Accessed 05-11-2025]},
}

@misc{blogReviewZeroday,
	author = {},
	title = {{A} review of zero-day in-the-wild exploits in 2023 --- blog.google},
	howpublished = {\url{https://blog.google/technology/safety-security/a-review-of-zero-day-in-the-wild-exploits-in-2023}},
	year = {2024},
}

@misc{cve,
	author = {},
	title = {Metrics | CVE},
	howpublished = {\url{https://www.cve.org/About/Metrics}},
	year = {},
	note = {[Accessed 05-11-2025]},
}

@article{anwar2020measuring,
  title={Measuring the Cost of Software Vulnerabilities.},
  author={Anwar, Afsah and Khormali, Aminollah and Alasmary, Hisham and Choi, Sung J and Salem, Saeed and Mohaisen, David and others},
  journal={EAI Endorsed Transactions on Security \& Safety},
  volume={7},
  number={23},
  year={2020}
}

@article{iannone2022secret,
  title={The secret life of software vulnerabilities: A large-scale empirical study},
  author={Iannone, Emanuele and Guadagni, Roberta and Ferrucci, Filomena and De Lucia, Andrea and Palomba, Fabio},
  journal={IEEE Transactions on Software Engineering},
  volume={49},
  number={1},
  pages={44--63},
  year={2022},
  publisher={IEEE}
}

@misc{cybersecurityventuresCybercrimeCost,
	author = {Cybercrime Magazine},
	title = {{C}ybercrime {T}o {C}ost {T}he {W}orld \$10.5 {T}rillion {A}nnually {B}y 2025},
	howpublished = {\url{https://cybersecurityventures.com/hackerpocalypse-cybercrime-report-2016}},
	year = {2020},
	note = {[Accessed 05-11-2025]},
}

@article{manes2019fuzzingSurvey,
  title={The art, science, and engineering of fuzzing: A survey},
  author={Man{\`e}s, Valentin JM and Han, HyungSeok and Han, Choongwoo and Cha, Sang Kil and Egele, Manuel and Schwartz, Edward J and Woo, Maverick},
  journal={IEEE Transactions on Software Engineering},
  volume={47},
  number={11},
  pages={2312--2331},
  year={2019},
  publisher={IEEE}
}

@article{serebryany2017oss,
  title={$\{$OSS-Fuzz$\}$-Google's continuous fuzzing service for open source software},
  author={Serebryany, Kostya},
  year={2017}
}

@article{jimenez2023swebench,
  title={Swe-bench: Can language models resolve real-world github issues?},
  author={Jimenez, Carlos E and Yang, John and Wettig, Alexander and Yao, Shunyu and Pei, Kexin and Press, Ofir and Narasimhan, Karthik},
  journal={arXiv preprint arXiv:2310.06770},
  year={2023}
}

@inproceedings{huang2024ntr,
  title={Template-guided program repair in the era of large language models},
  author={Huang, Kai and Zhang, Jian and Meng, Xiangxin and Liu, Yang},
  booktitle={Proceedings of the 47th International Conference on Software Engineering, ICSE},
  pages={367--379},
  year={2024}
}

@inproceedings{zhou2024vulMaster,
  title={Out of sight, out of mind: Better automatic vulnerability repair by broadening input ranges and sources},
  author={Zhou, Xin and Kim, Kisub and Xu, Bowen and Han, DongGyun and Lo, David},
  booktitle={Proceedings of the IEEE/ACM 46th International Conference on Software Engineering},
  pages={1--13},
  year={2024}
}

@inproceedings{liu2019tbar,
  title={TBar: Revisiting template-based automated program repair},
  author={Liu, Kui and Koyuncu, Anil and Kim, Dongsun and Bissyande, Tegawende F},
  booktitle={Proceedings of the 28th ACM SIGSOFT international symposium on software testing and analysis},
  pages={31--42},
  year={2019}
}

@inproceedings{yin2024thinkrepair,
  title={Thinkrepair: Self-directed automated program repair},
  author={Yin, Xin and Ni, Chao and Wang, Shaohua and Li, Zhenhao and Zeng, Limin and Yang, Xiaohu},
  booktitle={Proceedings of the 33rd ACM SIGSOFT International Symposium on Software Testing and Analysis},
  pages={1274--1286},
  year={2024}
}

@article{le2012GenProg,
  title={Genprog: A generic method for automatic software repair},
  author={Le Goues, Claire and Nguyen, ThanhVu and Forrest, Stephanie and Weimer, Westley},
  journal={Ieee transactions on software engineering},
  volume={38},
  number={1},
  pages={54--72},
  year={2011},
  publisher={IEEE}
}

@inproceedings{fu2022vulrepair,
  title={VulRepair: a T5-based automated software vulnerability repair},
  author={Fu, Michael and Tantithamthavorn, Chakkrit and Le, Trung and Nguyen, Van and Phung, Dinh},
  booktitle={Proceedings of the 30th ACM joint european software engineering conference and symposium on the foundations of software engineering},
  pages={935--947},
  year={2022}
}

@article{Renaud2015SpoonLibrary,
  TITLE = "{Spoon: A Library for Implementing Analyses and Transformations of Java Source Code}",
  AUTHOR = {Pawlak, Renaud and Monperrus, Martin and Petitprez, Nicolas and Noguera, Carlos and Seinturier, Lionel},
  JOURNAL = "{Software: Practice and Experience}",
  PUBLISHER = "{Wiley-Blackwell}",
  PAGES = {1155-1179},
  VOLUME = {46},
  URL = {https://hal.archives-ouvertes.fr/hal-01078532/document},
  YEAR = {2015},
  doi = {10.1002/spe.2346},
}

@article{liu2024deepseekv3,
  title={Deepseek-v3 technical report},
  author={Liu, Aixin and Feng, Bei and Xue, Bing and Wang, Bingxuan and Wu, Bochao and Lu, Chengda and Zhao, Chenggang and Deng, Chengqi and Zhang, Chenyu and Ruan, Chong and others},
  journal={arXiv preprint arXiv:2412.19437},
  year={2024}
}

@article{bouzenia2024repairagent,
  title={Repairagent: An autonomous, llm-based agent for program repair},
  author={Bouzenia, Islem and Devanbu, Premkumar and Pradel, Michael},
  journal={arXiv preprint arXiv:2403.17134},
  year={2024}
}

@misc{openaiTextEmbedding3Models,
	author = {OpenAI},
	title = {New embedding models and API updates},
	howpublished = {\url{https://openai.com/index/new-embedding-models-and-api-updates}},
	year = {},
	note = {[Accessed 25-05-2025]},
}

@inproceedings{zhang2024autocoderover,
  title={Autocoderover: Autonomous program improvement},
  author={Zhang, Yuntong and Ruan, Haifeng and Fan, Zhiyu and Roychoudhury, Abhik},
  booktitle={Proceedings of the 33rd ACM SIGSOFT International Symposium on Software Testing and Analysis},
  pages={1592--1604},
  year={2024}
}

@article{zhang2025ReinFix,
  title={Repair Ingredients Are All You Need: Improving Large Language Model-Based Program Repair via Repair Ingredients Search},
  author={Zhang, Jiayi and Huang, Kai and Zhang, Jian and Liu, Yang and Chen, Chunyang},
  journal={arXiv preprint arXiv:2506.23100},
  year={2025}
}

@misc{antoniades2025swesearchenhancingsoftwareagents,
      title={SWE-Search: Enhancing Software Agents with Monte Carlo Tree Search and Iterative Refinement}, 
      author={Antonis Antoniades and Albert Örwall and Kexun Zhang and Yuxi Xie and Anirudh Goyal and William Wang},
      year={2025},
      eprint={2410.20285},
      archivePrefix={arXiv},
      primaryClass={cs.AI},
      url={https://arxiv.org/abs/2410.20285}, 
}

@ARTICLE{chen2023vrepair,
  author={Chen, Zimin and Kommrusch, Steve and Monperrus, Martin},
  journal={IEEE Transactions on Software Engineering}, 
  title={Neural Transfer Learning for Repairing Security Vulnerabilities in C Code}, 
  year={2023},
  volume={49},
  number={1},
  pages={147-165},
  doi={10.1109/TSE.2022.3147265}}

@inproceedings{nong2025appatch,
  title={$\{$APPATCH$\}$: Automated Adaptive Prompting Large Language Models for $\{$Real-World$\}$ Software Vulnerability Patching},
  author={Nong, Yu and Yang, Haoran and Cheng, Long and Hu, Hongxin and Cai, Haipeng},
  booktitle={34th USENIX Security Symposium (USENIX Security 25)},
  pages={4481--4500},
  year={2025}
}

@Inbook{SafetyProperties,
author="B{\'e}rard, B{\'e}atrice
and Bidoit, Michel
and Finkel, Alain
and Laroussinie, Fran{\c{c}}ois
and Petit, Antoine
and Petrucci, Laure
and Schnoebelen, Philippe
and Mckenzie, Pierre",
title="Safety Properties",
bookTitle="Systems and Software Verification: Model-Checking Techniques and Tools",
year="2001",
publisher="Springer Berlin Heidelberg",
address="Berlin, Heidelberg",
pages="83--89",
isbn="978-3-662-04558-9",
doi="10.1007/978-3-662-04558-9_7",
url="https://doi.org/10.1007/978-3-662-04558-9_7"
}

@inproceedings{propertyGPT,
  author       = {Ye Liu and
                  Yue Xue and
                  Daoyuan Wu and
                  Yuqiang Sun and
                  Yi Li and
                  Miaolei Shi and
                  Yang Liu},
  title        = {PropertyGPT: LLM-driven Formal Verification of Smart Contracts through
                  Retrieval-Augmented Property Generation},
  booktitle    = {32nd Annual Network and Distributed System Security Symposium, {NDSS}
                  2025, San Diego, California, USA, February 24-28, 2025},
  publisher    = {The Internet Society},
  year         = {2025},
  bibsource    = {dblp computer science bibliography, https://dblp.org}
}

@inproceedings{useSafePropToGenVulPatch,
  author       = {Zhen Huang and
                  David Lie and
                  Gang Tan and
                  Trent Jaeger},
  title        = {Using Safety Properties to Generate Vulnerability Patches},
  booktitle    = {2019 {IEEE} Symposium on Security and Privacy, {SP} 2019, San Francisco,
                  CA, USA, May 19-23, 2019},
  pages        = {539--554},
  publisher    = {{IEEE}},
  year         = {2019},
  url          = {https://doi.org/10.1109/SP.2019.00071},
  doi          = {10.1109/SP.2019.00071},
  bibsource    = {dblp computer science bibliography, https://dblp.org}
}

@misc{githubDeepSeekV32ExpDeepSeek_V3_2pdfMain,
	author = {DeepSeek AI},
	title = {DeepSeek-V3.2-Exp: Boosting Long-Context Efficiency
with DeepSeek Sparse Attention},
	howpublished = {\url{https://github.com/deepseek-ai/DeepSeek-V3.2-Exp/blob/main/DeepSeek_V3_2.pdf}},
	year = {2025},
	note = {[Accessed 06-11-2025]},
}

@article{li2025sokAVR,
  title={Sok: Towards effective automated vulnerability repair},
  author={Li, Ying and Shezan, Faysal Hossain and Wei, Bomin and Wang, Gang and Tian, Yuan},
  journal={arXiv preprint arXiv:2501.18820},
  year={2025}
}

@inproceedings{pearce2023examiningZeroShotVulRepairWithLLMs,
  title={Examining zero-shot vulnerability repair with large language models},
  author={Pearce, Hammond and Tan, Benjamin and Ahmad, Baleegh and Karri, Ramesh and Dolan-Gavitt, Brendan},
  booktitle={2023 IEEE Symposium on Security and Privacy (SP)},
  pages={2339--2356},
  year={2023},
  organization={IEEE}
}

@misc{githubGitHubEclipsejdtlseclipsejdtls,
	author = {},
	title = {Eclipse JDT Language Server},
	howpublished = {\url{https://github.com/eclipse-jdtls/eclipse.jdt.ls}},
	year = {},
	note = {[Accessed 09-11-2025]},
}

@misc{microsoftOfficialPageLSP,
	author = {Microsoft},
	title = {{L}anguage {S}erver {P}rotocol},
	howpublished = {\url{https://microsoft.github.io/language-server-protocol}},
	year = {},
}

@misc{githubGitHubVndeellmsandbox,
	author = {Duy Huynh},
	title = {{G}it{H}ub - vndee/llm-sandbox: {L}ightweight and portable {L}{L}{M} sandbox runtime (code interpreter) {P}ython library.},
	howpublished = {\url{https://github.com/vndee/llm-sandbox}},
	year = {},
}

@misc{llvmClangFormatx2014,
	author = {Clang Contributors},
	title = {{C}lang{F}ormat},
	howpublished = {\url{https://clang.llvm.org/docs/ClangFormat.html}},
	year = {},
}

@misc{githubGitHubJlefflersccsnapshots,
	author = {Jonathan Leffler},
	title = {{G}it{H}ub - jleffler/scc-snapshots: {S}{C}{C}: {S}trip {C} {C}omments},
	howpublished = {\url{https://github.com/jleffler/scc-snapshots}},
	year = {},
}

@misc{llvmLibclangTutorial,
	author = {Clang Contributors},
	title = {{L}ibclang},
	howpublished = {\url{https://clang.llvm.org/docs/LibClang.html}},
	year = {},
}

@misc{llvmWhatClangd,
	author = {Clangd Contributors},
	title = {{W}hat is clangd?},
	howpublished = {\url{https://clangd.llvm.org}},
	year = {},
}

@misc{githubGitHubSWEagentSWEReX,
	author = {SWE-ReX Contributors},
	title = {{G}it{H}ub - {S}{W}{E}-agent/{S}{W}{E}-{R}e{X}: {S}andboxed code execution for {A}{I} agents},
	howpublished = {\url{https://github.com/SWE-agent/SWE-ReX}},
	year = {},
}

@misc{langchainLangChain,
	author = {LangChain Team},
	title = {{L}ang{C}hain},
	howpublished = {\url{https://www.langchain.com}},
	year = {},
}

@article{tao2024magis,
  title={Magis: Llm-based multi-agent framework for github issue resolution},
  author={Tao, Wei and Zhou, Yucheng and Wang, Yanlin and Zhang, Wenqiang and Zhang, Hongyu and Cheng, Yu},
  journal={Advances in Neural Information Processing Systems},
  volume={37},
  pages={51963--51993},
  year={2024}
}

@article{meng2024empiricalStudyOnLLMBasedBugFixing,
  title={An empirical study on llm-based agents for automated bug fixing},
  author={Meng, Xiangxin and Ma, Zexiong and Gao, Pengfei and Peng, Chao},
  journal={arXiv preprint arXiv:2411.10213},
  year={2024}
}

@article{chenYeHe2025prometheus,
  title={Prometheus: Unified Knowledge Graphs for Issue Resolution in Multilingual Codebases},
  author={Chen, Zimin and Pan, Yue and Lu, Siyu and Xu, Jiayi and Goues, Claire Le and Monperrus, Martin and Ye, He},
  journal={arXiv preprint arXiv:2507.19942},
  year={2025}
}

@article{yang2024sweBenchMultimodal,
  title={Swe-bench multimodal: Do ai systems generalize to visual software domains?},
  author={Yang, John and Jimenez, Carlos E and Zhang, Alex L and Lieret, Kilian and Yang, Joyce and Wu, Xindi and Press, Ori and Muennighoff, Niklas and Synnaeve, Gabriel and Narasimhan, Karthik R and others},
  journal={arXiv preprint arXiv:2410.03859},
  year={2024}
}

@article{rahardja2025canAgentsFixAgentIssues,
  title={Can Agents Fix Agent Issues?},
  author={Rahardja, Alfin Wijaya and Liu, Junwei and Chen, Weitong and Chen, Zhenpeng and Lou, Yiling},
  journal={arXiv preprint arXiv:2505.20749},
  year={2025}
}

@misc{githubIssueNJS482,
	author = {},
	title = {{S}{E}{G}{V} njs\_string.c:2535:18 in njs\_string\_offset · {I}ssue \#482 · nginx/njs},
	howpublished = {\url{https://github.com/nginx/njs/issues/482}},
	year = {2022},
	note = {[Accessed 13-11-2025]},
}

@misc{githubsecbenchSWEagent,
	author = {},
	title = {SEC-bench/SWE-agent},
	howpublished = {\url{https://github.com/SEC-bench/SWE-agent}},
	year = {2025},
	note = {[Accessed 13-11-2025]},
}

@misc{githubsecbenchOpenHands,
	author = {},
	title = {SEC-bench/OpenHands},
	howpublished = {\url{https://github.com/SEC-bench/OpenHands}},
	year = {2025},
	note = {[Accessed 13-11-2025]},
}

@misc{githubsecbenchAider,
	author = {},
	title = {SEC-bench/aider},
	howpublished = {\url{https://github.com/SEC-bench/aider}},
	year = {2025},
	note = {[Accessed 13-11-2025]},
}

@misc{secbenchSECbenchLeaderboard,
	author = {},
	title = {{S}{E}{C}-bench Leaderboard},
	howpublished = {\url{https://sec-bench.github.io}},
	year = {},
	note = {[Accessed 14-11-2025]},
}

\end{document}